\documentclass[journal]{IEEEtran}

\usepackage{url}
\usepackage{booktabs}
\usepackage{longtable}
\usepackage{xcolor}
\usepackage{hyperref} 
\usepackage{tabularx}
\usepackage{graphicx} 
\usepackage{pdflscape}
\usepackage{enumitem}

\begin{document}
\title{A comprehensive overview of the protocols associated with Intelligent Transportation Systems}

\author{Jonas~Vogt,~\IEEEmembership{Member,~IEEE}
\thanks{J. Vogt is with ITS research group at htw saar Saarbrücken, Germany, and with Institute for Wireless Communication and Navigation at RPTU Kaiserslautern, Germany, e-mail: jonas.vogt@htwsaar.de}
}

\markboth{RPTU Kaiserlautern - htw saar - Research White Paper - June 2024}%
{Vogt: A comprehensive overview of the protocols associated with Intelligent Transportation Systems}


\maketitle

\begin{abstract}
This white paper offers a comprehensive overview of the protocols utilized in the field of intelligent transportation systems (ITS).
The paper presents a comprehensive overview of protocols from all layers of the Open System Interconnection (OSI) model.
The protocols encompass a range of communication technologies, including ad-hoc, mobile broadband, and fiber-optic cable.
All of these protocols are utilized in the field of intelligent transportation systems (ITS) and must exchange information at some point.
The current development leads to a much tighter connection between ecosystems that were previously isolated or not connected.
This situation necessitates a comprehensive understanding of the protocols at all layers to ensure the establishment of safe and secure applications within the field of intelligent transportation systems (ITS).
\end{abstract}


\section{Introduction} 
\label{Chapter_ITSProtocols-Introduction}

\IEEEPARstart{T}{he} field of intelligent transport systems (ITS) encompasses the entirety of the transportation sector, encompassing all modes of transportation, including land, water, and air. 
For the purpose of this analysis, only those protocols that are pertinent to the communication of street vehicles will be considered.
The following analysis does not consider the areas of rail-based transport, water, and air.
Some protocols could be applied to these areas because they are very generic, even though some are not designed with these specific areas in mind.
Additionally, these protocols are highly specialized and have a limited scope, rendering them unsuitable that for other purposes.
This can have different reasons. 
For example, some protocols are based on specific traffic rules or laws in a country (see chapter \ref{Chapter_ITSProtocols-TP-PS-RiLSA}), while others are developed by a group of manufacturers (see chapter \ref{Chapter_ITSProtocols-TP-PS-OCIT}).

This paper introduces a wide range of protocols. 
This introduction is limited to the essential protocols and related specifications and does not claim to be exhaustive.
The protocols are presented in varying degree of detail.
The reader is invited to discover for oneself the protocols that are available free of charge, which are only briefly introduced here.
Protocol descriptions that require financial or personal data transactions are presented in greater detail.

The paper is divided into six sections. 
Section \ref{Chapter_ITSProtocols-related_work}, entitled "Related Work," presents an overview of previous research in the field. 
Section \ref{Chapter_ITSProtocols-TrafficProtocols}, titled "Traffic-Related Application Protocols," introduces various protocols designed for use in transportation and traffic management. 
Section \ref{Chapter_ITSProtocols-CommunicationProtocols}, "Communication-Related Protocols," describes communication protocols used in various contexts, including those related to intelligent transportation systems (ITS). 
Section \ref{Chapter_ITSProtocols-ConnArchDataExchange}, "Architectures and Data Exchange Platforms," provides an overview of various architectures and data exchange platforms used in ITS. 
Section \ref{Appendix_PE}, "Detailed Traffic-Related Specifications," presents detailed interpretations of specifications related to traffic management. 
The final section \ref{Chapter_ITSProtocols-outlook}, "Outlook," presents a prospective view of future developments in the field.

\section{Related work} 
\label{Chapter_ITSProtocols-related_work}

The number of relevant standards changes frequently. 
New standards are published, old specifications are withdrawn, and current ones are updated. 
Standards organizations have internal and external mailing lists and Web sites to keep track of standards.
There are books and book chapters (e.g., \cite{ITSStandards2008}, \cite{Ernst2021}) and papers that provide an overview of standardization activities.

In \cite{Festag2009}, the authors describe the early European standardization processes and goals in 2009.
In \cite{Festag2014}, \cite{Kosch2009} and \cite{Sousa2017} the European ITS Station Reference Architecture is described.
It focuses on the cooperative ITS (C-ITS) part for direct communication.
There are papers describing the status of ITS standardization and deployment in different countries or regions. 
The situation for Poland is described in \cite{Macioszek2014}.
For Asia, \cite{Feng2017} describes the standardization in China, \cite{Lim2012} in Korea, and \cite{Sugimoto1999} presents the plans for ITS in Japan.
The special situation of ITS deployment in developing countries is described in \cite{Padmadas2010} (with the example of license plate recognition) and \cite{Djalalov2013} (the role of standardization in general).
Security aspects related to the architecture are described in \cite{Lonc2016} and \cite{Hamid2015}.
Some identity and credential management requirements are presented in \cite{Khodaei2015}.
The data distribution and knowledge systems are presented in \cite{Pri2021}.
ITS architecture requirements for information distribution based on application viewpoints are given in \cite{Ren2001}.
Both are also presented in the results of research projects such as $sim^{TD}$ \cite{Stuebing2010} or C-MOBilE \cite{Lu2018}.
Specific communication-related standardization issues are addressed in \cite{Nsonga2016} (EV IEC 61850 and WAVE) and \cite{Zeadally2020} (IEEE 802.11bd and 5G NR V2X).
Special attention must be paid to functional safety for ITS interoperability \cite{Mariani2021}.
These papers provide a broad overview of the field, but none of the literature provides insight into the standardization activities, the standardization organizations, and the relationship between the organizations.
There has also been no review of ETSI ITS Release 1.

The various standards organizations provide overviews of their current standardization activities.
For example, the \textit{Comite Europeen de Normalisation} (CEN) has three web pages for this purpose.
The website of Technical Committee (TC) 278 ITS Standardization\footnote{CEN technical committee (TC) 278 ITS standardization website. [Online]. Available:  \url{https://www.itsstandards.eu/}} describes the work and activities of the TC. 
The CEN website \footnote{CEN standards search website. [Online]. Available:  \url{https://standards.cencenelec.eu/dyn/www/f?p=205:105:0:::::}} allows searching for all standards related to ITS.
TC 278 provides the \textit{EU-ICIP GUIDE} (ICIP = ITS Communication and Information Protocols)\footnote{ITS Communication and Information Protocols. [Online]. Available:  \url{https://www.mobilityits.eu/home}}. 
The website provides information about ITS in general and standardization. 

The \textit{European Telecommunications Standards Institute} (ETSI) provides general information about ITS standardization on the page of the TC ITS\footnote{ETSI Technical Committee (TC) Intelligent Transport Systems (ITS). [Online]. Available: \url{https://www.etsi.org/committee/1402-its}}.
ETSI publishes all documents free of charge on the Internet\footnote{ETSI standards overview. [Online]. Available: \url{https://www.etsi.org/standards}}.
Current and previous versions can be downloaded directly\footnote{ETSI standards download. [Online]. Available: \url{https://www.etsi.org/deliver/etsi_ts}}.

The \textit{International Standardization Organization} (ISO) TC 204 is responsible for ITS-related specifications\footnote{ISO Technical Committee (TC) 204. [Online]. Available: 
\url{https://www.iso.org/committee/54706.html}}.
A list of all specifications is published by TC 204\footnote{ISO TC 204 specifications. [Online]. Available: \url{https://www.iso.org/committee/54706/x/catalogue/p/1/u/0/w/0/d/0}}.

A summary of many standardization documents from various SDOs is available from the \textit{International Telecommunication Union} (ITU)\footnote{ITU ITS specification overview. [Online]. Available: \url{https://www.itu.int/net4/ITU-T/landscape\#?topic=0.131&workgroup=1}}. 
The site allows you to search for documents from 16 different organizations, and the database contains 1,265 files.

\section{Traffic-realted applications protocols} 
\label{Chapter_ITSProtocols-TrafficProtocols}

In the field of transportation, a multitude of regulations and statutes exist (e.g., StVo). 
These legal frameworks establish the parameters within which traffic protocols are implemented.

This section is divided into three sections. 
Subsection \ref{Chapter_ITSProtocols-TP-TPS} describes protocols related to infrastructure-based traffic-related communication.
This includes protocols for communication between back-end systems and the traffic infrastructure on the roadside and between back-end systems and the traffic participants.
Subsection \ref{Chapter_ITSProtocols-TP-ITSApp} describes the protocols for direct communication between the traffic participants and between the traffic participants and the infrastructure.
Additionally, some framework specifications are introduced to provide some context.
The third subsection, \ref{Chapter_ITSProtocols-TP-ITSApp-secp}, adresses some security and privacy-related protocols and context information.

\subsection{Infrastructure traffic protocol standards}
\label{Chapter_ITSProtocols-TP-TPS}
This section describes the protocols utilized to disseminate pertinent information from central traffic management centers (TMCs) to traffic infrastructure situated on the roadside or between TMCs or other back-end systems.
It should be noted that the majority of the protocols described in this section must be procured.
Those protocols are described in greated detail below.

\subsubsection{DATEX2}
\label{Chapter_ITSProtocols-TP-PS-DATEX2}
DATEX II \cite{CEN16157-1} is an European specification for data exchange between different TMCs and TMCs and other ITS service providers. 
The primary motivation for its creation was the establishment of a pan-European mobility space. 
An introduction to DATEX II can be found on \url{https://www.datex2.eu/} and \url{https://docs.datex2.eu/}.  
Unfortunately, the site was not updated since 2021.
Consequently, some of the more recent specifications are missing.
DATEX II defines data formats based on UML and XML for all kinds of traffic-related information.

Currently, it comprises the following twelfth parts:
\begin{itemize}
	\item \textbf{Part 1: Context and framework} describes the manner in which UML should be employed to extend DATEX II.
	\item \textbf{Part 2: Location referencing} defines the system that may be used for referencing locations.
	\item \textbf{Part 3: Situation Publication} defines the data elements, roles, and relationships for traffic and travel information.
	\item \textbf{Part 4: Variable Message Sign (VMS)} specifies how VMS's graphic and textual content can be described. It does not support the control or configuration of VMS.
	\item \textbf{Part 5: Measured and elaborated data publications} defines data formats for values measured or calculated by an infrastructure system.
	\item \textbf{Part 6: Parking Publications} specifies data elements related to parking information, including location, special spot types, and rates.
	\item \textbf{Part 7: Common data elements} defines data elements used in multiple specifications, including fundamental data types, vehicle-related data types, or time-related data types.
	\item \textbf{Part 8: Traffic management publications and extensions dedicated to the urban environment} describes an extension for urban scenarios (e.g., lanes, information to users, work zones, etc.) and an extension for traffic management plans and traffic routing.
	\item \textbf{Part 9: Traffic signal management publications dedicated to the urban environment} defines the XML representation for MAPEM and SPATEM as described in sections \ref{Chapter_ITSProtocols-TP-ITSApp-MAP} and \ref{Chapter_ITSProtocols-TP-ITSApp-SPaT} respectively.
	\item \textbf{Part 10: Energy infrastructure publications} defines data elements for all kinds of charging (electricity) and refueling (gas, petrol, hydrogen, ...) infrastructure, including status, price, or location. 
	\item \textbf{Part 11: Publication of machine-interpretable traffic regulations} specifies data elements concerning traffic regulation published by road authorities. These data elements are not intended for end users but rather for ITS service providers.
	\item \textbf{Part 12: Facility related publications} defines generic data elements to be used by other specifications related to dimensions, organizational information, rates, opening hours, and other supplementary systems.
\end{itemize}

In theory, DATEX II may be transported via every transport or application layer protocol.
To reduce the number of options, DATEX II defines a concept of a so-called exchange specification.
The current version is \textit{Exchange 2020}.
The most straightforward information-gathering form is based on HTTP/get.
However, this procedure was not deemed reliable enough, so the concept of \textit{Platform Independent Model} (PIM) and \textit{Platform Specific Model} (PSM) was introduced.
PIM is defined in ISO TS 19468 \cite{TS19468}, while PSM is specified in \cite{TS14827-4}.
PIM and PSM are connected via the \textit{Functional Exchange Profile} (FEP) and \textit{Exchange Pattern} (EP).
Currently, the only PSM available is SOAP\footnote{W3C SOAP specification \url{https://www.w3.org/TR/2000/NOTE-SOAP-20000508/}}. 
The most common combination is SOAP using HTTP and TCP.
Both the public/subscribe and the request/response pattern may be used.
The data may be obtained in its entirely (a snapshot) or in part (simple).
It may be obtained in a stateful or stateless configuration.
For collaborative ITS services (CIS), service requests and feedback are defined.
In addition to application data, communication management information (e.g., availability, time, etc.) may be transmitted.

The existing XML/XSD specifications are extensive.
Therefore, profiles for every configuration were defined to make the information more easily accessible and to reduce computation time.

It should be noted that the data exchange with DATEX II is clear and the data encoding unambiguous.
However, systems using different coding may interpret XML data differently.
To mitigate this possibility, ASN.1 is mentioned as a potential alternative to XML.
However, no further standardization effort is made in this direction.

\subsubsection[NTCIP]{National Transportation Communications for Intelligent Transportation System Protocol (NTCIP)}
\label{Chapter_ITSProtocols-TP-PS-NTCIP}
In 1996, the 'National Transportation Communications for Intelligent Transportation System Protocol' (NTCIP) was introduced with the objective of promoting interoperability and interchangeability in the ITS domain.
NTCIP is a series of specifications for the communication between TMCs (referred to as C2C, which should not be confused with the car-to-car communication, which uses the same acronym) and the communication between TMCs and the field equipment (referred to as C2F).
The introduction of this protocol enables the exchange of systems, products, and vendors over time without the need for the implementation of new interfaces or vendor-specific solutions. 
This is achieved through the provision of a vendor-independent communication and data setup.
The NTCIP is published by the American Association of State Highway and Transportation Officials (AASHTO), the Institute of Transportation Engineers (ITE), and the National Electrical Manufacturers Association (NEMA).
An overview of the system design is presented in \cite{NTCIP9001}.
The specification numbers are four digits long and grouped by their numbers.
Specifications commencing with digit 1 are designated as application specifications, digit 2 signifies communication, digit 8 pertains to management, and digit 9 is reserved for testing and general information.
The solution is widely utilized in the United States but not in Europe.

The NTCIP \textit{framework} is categorized into five levels: $i$ information, $ii$ application, $iii$ transport, $iv$ subnetwork, and $v$ plant.
Different protocols and technologies operationalize the distinct levels.
The levels correspond to the OSI layers.
Only the combinations shown in Figure 4 of \cite[p. 12]{NTCIP9001} are permitted.
At the plant level, different access technologies are identified.
These include dial-up telco, fiber, coax, wireless, twisted pair, and leased line.
It should be noted that the plant level is the only level not included in the NTCIP specifications.
In the following sections, the numbers attached to certain specifications indicate the profile specified in NTCIP regarding this technology.

At the subnetwork level, the following protocols are employed: Point-to-Point Protocol (PPP, RFC 1661, 2103), Ethernet (\cite{IEEE802.3}, 2104), and Point-to-Multipoint Protocol (PMP, using High-Level Data Link Control (HDLC) via RS232, 2101+2102).
The transport layer employs the following combinations: TCP/IP (2202), UDP/IP (2202), and T2/NULL (own protocol, 2201).
At the application layer, C2C XML (utilizing XML or SOAP, 2306), DATEX (see chapter \ref{Chapter_ITSProtocols-TP-PS-DATEX2}, employing ASN.1, 2304), File Transfer Protocol (FTP, RFC 959, 2303), Trivial File Transfer Protocol (TFTP, RFC 1350, 2302), Simple Network Management Protocol (SNMP, RFC 1157, 2301) and Simple Transportation Management Protocol (STMP, 2301).
C2C messages, files, data objects, and dynamic objects are specified at the application layer.
They comprise the C2C data dictionaries (from different standardization organizations) and the NTCIP data dictionaries (12xx).
On top of the functional area, data dictionaries are used.
A specification does not represent the latter.

NTCIP may utilizes three communications patterns.
\begin{itemize}
	\item Request-Response: Synchronous communication.
	\item Dynamic Objects: Subscription with asynchronous information delivery on change.
	\item Traps: Field devices send information when they occur without a subscription to the center.
\end{itemize}

\subsubsection[OCIT]{Open Communication Interfaces for Road Traffic Control Systems (OCIT)}
\label{Chapter_ITSProtocols-TP-PS-OCIT}
The OCIT (Open Communication Interfaces for Road Traffic Control Systems) is standardized protocol developed by the industry consortia of companies (OCIT Developer Group\footnote{OCIT developer Group website: \url{https://www.ocit.org/en/}}, ODG) that produce traffic regulation systems for the roadside.
In general, OCIT comprises two communication paradigms: OCIT-C (Center to Center) for the communication between TMCs (central traffic management, traffic light management, parking management, construction management, internet data presentation, and interface to other protocols like TLS, VDV (Verband Deutscher Verkehrsunternehmen, e.g., public transport prioritization), and Datex II) and OCIT-O (Outstation) for the communication between the TMC and the traffic light.
Furthermore, the OCIT-LED is designated for communication between the traffic light controller and light-emitting diode (LED) based traffic lights.
An overview of those specifications can be found in \cite[p. 7]{OCIToSystem}.
The following outlines the principles for the outstation and the center specification.

\paragraph{OCIT-O}
\label{Chapter_ITSProtocols-TP-PS-OCIT-OCIT-O}
The current version of OCIT-O is 3.0.
The OCIT-O has been in a beta status since 2018, and the website indicates that changes to the specification may be made. 
However, there have been no changes since the initial release.

The protocol for OCIT-O is specified in \cite{OCIToProtocol}.
Its design is intended to provide communication over low-speed networks, such as dial-in lines or GSM connections, as well as high-speed networks, such as fiber optic networks. 
Therefore, it should be highly resource efficient.
For the different technologies and networks (private or public), OCIT-O has four profile documents for communication.
Profile one is designed for point-to-point connections in circuit-switched networks.
Profile two is intended for dial-in connections and GSM.
Profile three specifies the parameters for Ethernet with DHCP.
Finally, profile four provides information for VPN with OpenVPN\footnote{OpenVPN is an open source VPN solution via secured transport layer Security (TLS) connections: OpenVPN website: \url{https://openvpn.net/}} and certificate management.
The first two OSI layers can be freely selected for OCIT-O.
On the network layer, only IP is specified. 
No version is specified for IP.
UDP and TCP are possible on the transport layer.
However, UDP is no longer recommended.
Above the transport layer, OCIT-O specifies the basic transport packet protocol layer (in German: Basis Transport Paket Protokoll Layer (BTPPL)).
BTPPL is specified by a table and byte-based representation of fields.
It uses the TCP port 3110 for priority messages.
The abbreviation is PNP.
The high priority (PHP) port is 2504.
Why these numbers were chosen is not specified\footnote{According to the Internet Assigned Numbers Authority (IANA). 
Port 2504 is used for the Windows NT load Balancing Service, and port 3110 is the Simulator control port.}\footnote{For more information on ports: \url{https://www.iana.org/assignments/service-names-port-numbers/service-names-port-numbers.xhtml}}.

In terms of security, SHA-1 is used for message integrity. 
It should be mentioned that the National Institute of Standards and Technology (NIST) has prohibited the usa of SHA-1 as of 2013.
However, a Hash-Based Message Authentication Code (HMAC) can be used for the time being.
The hash is usually calculated by hashing the concatenation of the secret key (extended with 0-byte padding to the block length of the hash function), the hash of the secret key, and the message itself.
In contrast, OCIT-O hashes the concatenation of the secret key (expanded to 512 bits), the message, and the compressed key.

Four data categories have been identified for OCIT-O: supply objects, central switch requests, messages and measurements, and V2X communication. 
The specification is based on XML and combines the operations get, update, create, and delete.

The V2X data set encompasses elements of the CAM message pertaining to vehicles, public transport prioritization, DENM, SPaT, and MAP.
OCIT-O is equipped with an extension for the communication with vehicles: OCIT-O car \cite{OCIToCar}. 
This specification facilitates communicate between a TMC and roadside stations.
It delineates the data to be exchanged between the RSU and the TMC.
At present, there is no direct communication between the traffic light controller and the RSU.
The CAM message can be employed for prioritization of public transportation. 
The data set includes a PublicTransportContainer, which contains information about passengers currently entering and leaving the vehicle.
Additionally, it contains information about the request for prioritization. 
The prioritization request includes the currently utilized R09.x protocol message, which is transmitted via analog signals and will be discontinued in a few years.
The R09.x protocol is specified in VDV 420\footnote{VDV is the Association of German Transport Operators (Verband Deutscher Verkehrsunternehmen, VDV). But the specification is used in several European countries. More information can be found at: \url{https://www.vdv.de/schriften---mitteilungen.en.aspx?id=acc1b4ae-c380-4fda-92ef-69950bd77f52&mode=detail}}.
In the present era, the European Telecommunications Standards Institute (ETSI) has deliniated the specifications for SRM and SSM, which are designed for this purpose.
In contrast to the R09.x protocol message, a technical feedback message is included.
In the absence of a more suitable solution, the driver will be informed via an additional light at the traffic light.
This is not the optimal approach for automated vehicles. 

\paragraph{OCIT-C}
\label{Chapter_ITSProtocols-TP-PS-OCIT-OCIT-C}
OCIT-C represents the communication framework for the connection between different TMCs.
The version 1.0 of OCIT-C is standardized in DIN VDE V 0832-601 and DIN VDE V 0832-602.
The former one defines the protocol and the data, while the latter is the schema definition based on XML.
The protocol of OCIT-C \cite{OCITcProtocol} is based on SOAP (no abbreviation) specified by W3C (the current version is 1.2 from the year 2007).
The web service is defined with the Web Services Description Language (WSDL).
SOAP is currently not used as frequently as it once was, but it has some advantages over other solutions.
However theses advantages are offset by additional overhead.

The OCIT-C protocol defines 15 data categories specified, including traffic messages, traffic data, parking, weather, camera, situation and strategy, maintenance, signs, public transport passenger information, traffic light information, traffic light raw data, V2X communication, traffic light basis data, binary container, and so-called project-related extensions. 

For security purposes, the username and password are utilized.
Theses credentials should be stored on both the server and the client, in conjunction with additional privileges. 
The combination is transmitted in plain text.
It relies on security mechanisms at the lower layers.
Since SOAP is based on Hypertext Transfer Protocol (HTTP), possible security measures utilize Transport Layer Security (TLS) and mutual certificate-based authentication.
Mutual trust should be based on mutual authentication. 
Therefore, a system based on username and password should be avoided.

\subsubsection[OCPP]{Open Charge Point Protocol (OCPP)}
\label{Chapter_ITSProtocols-TP-PS-OCPP}
The Open Charge Point Protocol (OCPP) was developed by the Open Charge Alliance for the connection between a charging station and the back-end systems \cite{OCPP-P1}. 
It complements the V2G protocol ISO 15118 family between the electric vehicle and the charging station described in section \ref{Chapter_ITSProtocols-TP-PS-V2G}.
It includes information for that protocol.
In its current version 2.0.1 from 2020, OCPP communicates via WebSocket \cite{RFC6455} with JavaScript Object Notation (JSON) \cite{RFC8259} formatted data \cite{OCPP-P2} .
Prior to version 2.0.0, also communication via SOAP (see section \ref{Chapter_ITSProtocols-CP-IP-SOAP}) was also possible.
In order to clarify the specification and reduce implementation overhead, SOAP was dropped, and only WebSockets are used.
OCPP enables bidirectional communication between the charging station and the back-end.
It can be used for management, configuration, accounting, and maintenance.
The protocol is divided into an information model for the data and a device model for the involved systems.
Based on its abstract device model \cite[p. 6f]{OCPP-P1}, OCPP can be used, in addition to charging stations, for communication with battery packs, transformers, and other electric-charging-related systems.

\subsubsection[RiLSA]{Guidelines for Light Signal Systems - Traffic Lights for Road Traffic (RiLSA)}
\label{Chapter_ITSProtocols-TP-PS-RiLSA}
The Guidelines for Light Signal Systems - Traffic Lights for Road Traffic \cite{Ril15} (original German title: Richtlinien für Lichtsignalanlagen (RiLSA) - Lichtzeichenanlagen für den Straßenverkehr) is the leading standard for traffic light systems in Germany. 
The RiLSA was developed by participants from road infrastructure providers and research for the German Federal Ministry for Transport and digital infrastructure (BMVI).
The ministry's name changed with the last election in 2021, and it is now called the German Federal Ministry for Digital and Transport (BMDV).

The RiLSA describes the steps to build and maintain a traffic light.
It outlines the requirements for the setup, the lights, and the corresponding rules to follow.
Additionally, the specifications include the requirements for the signal program for all participants (public transport, vehicles, pedestrians, etc.).
Theses requirements pertain to the minimum time intervals for green, yellow, and red, whihc are necessary for safe traffic.
The physical wiring is also included in the specification.
Finally, the control procedures, the technical execution, the approval process, and the quality management are laid out.
From a communication point of view, only soft requirements for exchanging information are described, but no details are provided.

\subsubsection[TLS]{Technical Delivery Conditions for Outstations (TLS)}
\label{Chapter_ITSProtocols-TP-PS-TLS}
The Technical Delivery Conditions for Outstations (TLS) \cite{TLS2012} (German title: Technische Lieferbedingungen für Streckenstationen) were initially specified in the year 2002.
The current version is from the year 2012.
The document's objective is to define physical specifications and interfaces, as well as logical communication interfaces.
This is done so that sensors on the roadside can provide information to TMC via outstations in a standardized and manufacturer-independent way on one side.
On the other side, outstations should be able to address and control gantries or other active traffic infrastructure (except traffic lights).
The outstations are, in both cases, a translator between a standardized communication to the TMC and an often proprietary communication to and from sensors or traffic infrastructure.

The design of an outstation should be based on a modular concept, allowing for the attachment of different sensors from different vendors.
The specification outlines the placement of physical interfaces.
It describes the data gathering, output, modules, physical framework conditions, interfaces, and test rules.

While communication is significant aspect of the specification, it is only one of many.
The specification defines different communication areas, which are referred to as local or island.
The network is divided into different subsystems, as follows: $\alpha$) the TMC, $\beta$) the subcenter (German: Unterzentrale), $\gamma$) the communication computer island bus (German Inselbus), and $\delta$) the outstation. 
In principle, the TMCs can communicate with each other and the subcenter.
The subcenter can communicate with the island bus or with the outstations.
Lastly, the island bus can communicate with the outstation.
The outstation is comprised of a controller that connects the I/O concentrates (referred to as a function groups), which provide access to sensors or actors.
A protocol is specified for communication at OSI layers 2 to 7 protocol.
In the 2012 version, IP is introduced as a new network layer (referred to as \textit{TLSoverIP}).
For IP, it is recommended to use TCP, and on public connections, a Secure Session Layer (SSL) should be used.
It is also worth noting that the newer IETF specified Transport Layer Security (TLS) should be used instead of the old Netscape SSL. 
The transition from SSL to TLS was made in the year 1996.
In the year 2012, the time of publishing the current version of technical delivery conditions for outstations, TLS version 1.2 was already valid for three years (published in 2008).
The idea for TLSoverIP is to encapsulate TLS-layer-2-datagrams inside of IP or TLS, respectively.

\subsubsection[TPEG2]{Transport Protocol Expert Group - Version 2 (TPEG2)}
\label{Chapter_ITSProtocols-TP-PS-TPEG2}
The Transport Protocol Expert Group (TPEG) is a language- and technology-independent toolbox for distributing mobility information for travelers.
The Intelligent Transport Systems - Traffic and Travel Information (TTI) via Transport Protocol Expert Group 2 (TPEG2) is an extension of the 1998 byte-oriented data stream standard TPEG based on the Unified Modeling Language (UML) and Extensible Markup Language (XML) 
The original TPEG standard was specified by the European Broadcast Union\footnote{EBU homepage: \url{https://www.ebu.ch/}} (EBU), whereas the current version is standardized by the Traveller Information Services Association \footnote{TISA website: \url{https://tisa.org/}} (TISA). 
The current TPEG2 standard supports both the XML and the binary formats. 
TISA specifies a toolkit for translating the UML-based description of the standard to both supported formats (compare \cite[p. vii]{TS21219-15}).

The section \ref{Appendix_PE-TP-TPEG2} presents an overview of the different sub-specification and an example dissection of the specification.

\subsubsection{ISO TS 19091}
\label{Chapter_ITSProtocols-TP-PS-TS19091}
The standard DIN CEN ISO TS19091 \cite{TS19091} Intelligent transport systems - Cooperative ITS - Using V2I and I2V communication for applications related to signalized intersections describes the use cases, functions, architecture, messages, and data elements related to direct communication in the context of traffic lights at crossroads.
The standard was developed and published by the three standardization institutes CEN, DIN, and ISO. 
It is based on ETSI and SAE standards and provides a baseline for North America, Japan, and Europe. 

The specification introduces three categories of use cases: $\alpha$) safety use cases (phase information, (possible) red light violation, etc.), $\beta$) mobility use cases (traffic efficiency, payment, etc.), and $\gamma$) priority use cases (public transport, emergency services, etc.).
It introduces the usage of the messages MAP (see \ref{Chapter_ITSProtocols-TP-ITSApp-MAP}), SRM (see \ref{Chapter_ITSProtocols-TP-ITSApp-SRM}), SPaT (see \ref{Chapter_ITSProtocols-TP-ITSApp-SPaT}), and SSM (see \ref{Chapter_ITSProtocols-TP-ITSApp-SSM}).
It is expected that the TLC will be connected to a RSU at the intersection.
The TLC will provide the controller and traffic-related information, and the RSU will generate the message (MAP, SRM, and SPaT).
The RSU will receive information from vehicles (CAM and SSM) and forward that information to the TLC.
The communication requirements describe how and when a message delivery should be expected.
The messages mentioned above are not specified in ISO TS 19091, but are part of SAE J2735 \cite{SAEJ2735}.
Therefore, only profiles of those messages for Europe, Japan, and North America are specified.
Additionally, regional extensions are mentioned.

More detailed information are provided in chapter \ref{Appendix_PE-TP-TS19091}.

\subsubsection{Vehicle to Grid Communication}
\label{Chapter_ITSProtocols-TP-PS-V2G}
The ISO 15118 vehicle to grid (V2G) family of standards defines the communication interface between charging infrastructure and electric vehicles. 
Standardization work commenced in 2010. 
The interface contains information relevant to bi-directional charging, including authentication, billing, charging procedure parameters, and additional value-added services. 
It also focuses on communication security and privacy. 
The standard comprises of the following parts:

\begin{itemize}
	\item ISO 15118-1: General information and use-case definition
	\item ISO 15118-2: Network and application protocol requirements
	\item ISO 15118-3: Physical and data link layer requirements
	\item ISO 15118-4: Network and application protocol conformance test
	\item ISO 15118-5: Physical and data link layer conformance test
	\item ISO 15118-6: General information and use-case definition for wireless communication (obsolete, included in ISO 15118-1)
	\item ISO 15118-7: Network and application protocol requirements for wireless communication (obsolete, included in ISO 15118-20)
	\item ISO 15118-8: Physical layer and data link layer requirements for wireless communication 
	\item ISO 15118-20: The second generation of network and application protocol requirements includes electric vehicles, bicycles, ships, and airplanes.
\end{itemize}

Further detailed information about ISO 15118 can be found in section \ref{Appendix_PE-TP-V2G}

\subsection{ITS Application Protocols}
\label{Chapter_ITSProtocols-TP-ITSApp}

The term 'application protocol' in this paper refers to all specifications above the transport layer of the Open Systems Interconnection (OSI) model.
This corresponds with the TCP/IP model.
This includes the facilities and the application layer for ITS.
First, a summary of ITS applications is provided.
Afterward, the support functionality is described.
Then the message service definitions are presented.
The last step presents the correlations between messages defined in different Standards Development Organizations (SDOs).

\subsubsection{ITS Application Overview}
\label{Chapter_ITSProtocols-TP-ITSAppOverview}

The Basic Set of Applications (BSA), as specified in \cite{TR102638}, describes the fundamental ITS applications involving vehicles based on V2V and V2I (or short V2X) communication.
These applications have already been standardized and are ready for deployment. 
The following list provides insight into applications that drove the development of the C-ITS communication system.
\begin{itemize}
	\item The Road Hazard Signaling (RHS) application describes the process of detecting a hazard, triggering the  dissemination of a message, and the subsequent reaction of a receiving station. This includes applications for collision avoidance and collision mitigation \cite{TS101539-1}. 
	\item The Intersection Collision Risk Warning (ICRW) application specifies the requirements for all kinds of possible collisions in the area of intersections between a vehicle and other traffic participants or obstacles \cite{TS101539-2}.
	\item Longitudinal Collision Risk Warning (LCRW) specifies the requirements for all kinds of possible collisions at the front or the rear side of a vehicle with other traffic participants or obstacles  \cite{TS101539-3}.
	\item Vulnerable Road User (VRU) applications combine all use cases with pedestrians, (motor)cyclists, animals, etc., where a dangerous situation, in most cases a collision can occur \cite{TR103300-1}, \cite{TS103300-2}, \cite{TS103300-3}.
	\item The following applications are related to intersections: Traffic Light Maneuver (TLM), Road and Lane Topology (RLT), and Traffic Light Control (TLC) \cite{TS103301}. 
	\item Infrastructure to Vehicle Information (IVI), In-Vehicle Service (IVS), and Contextual Speeds Indication (CSI) contain applications where signs and other traffic regulation information and informative data from the traffic infrastructure are transmitted to vehicles \cite{TS103301}, \cite{TS17425}, \cite{TS17426}.
	\item The GNSS Positioning Correction (GPC) applications provide infrastructure data to stations for a more precise positioning \cite{TS103301}.
\end{itemize}

As of 2024, extended and new applications are being developed, e.g., including scenarios with agriculture vehicles and motorcycles.
	
The EU project High Precision Positioning for Cooperative Intelligent Transportation Systems Applications (HIGHTS)\footnote{HIGHTS website: \url{https://ec.europa.eu/inea/en/horizon-2020/projects/h2020-transport/intelligent-transport-systems/hights}} compiled an ITS application overview based on ITS application requirements (\cite{EN17423}), the C2C-CC basic system profile (BSP) and the C-ROADS harmonized communication profile \cite{CROADSHCP} 
Additionally, the overview was compiled based on publications from the following organizations: European Commission C-ITS platform phase I report\footnote{EU website: \url{https://transport.ec.europa.eu/transport-themes/intelligent-transport-systems/cooperative-connected-and-automated-mobility-ccam_en}}, the Amsterdam Group\footnote{Amsterdam Group website: \url{https://amsterdamgroup.mett.nl/default.aspx}}, C2C-CC\footnote{C2C-CC website: \url{https://www.car-2-car.org}}, European Automotive and Telecoms Alliance (EATA)\footnote{EATA website: \url{https://eata.be/relevant-links/projects/}}, 5GAA\footnote{5GAA website: \url{https://5gaa.org}}, and European Automobile Manufacturers Association (for historic reasons ACEA)\footnote{ACEA website: \url{https://www.acea.be}}. Furthermore, the European projects C-Roads\footnote{C-Roads website: \url{https://www.c-roads.eu/platform.html}}, Interoperable Corridors (InterCor)\footnote{InterCor website: \url{https://intercor-project.eu}}, COoperative ITS DEployment Coordination Support (CODECS)\footnote{CODECS website: \url{https://www.codecs-project.eu/index.php?id=5}}, and other sources were consulted. 
The following list is initially based on a C2C-CC document\footnote{C2C-CC spectrum document: \href{https://www.car-2-car.org/fileadmin/documents/General\_Documents/C2CCC\_TR\_2050\_Spectrum\_Needs.pdf}{https://www.car-2-car.org/fileadmin/documents/General\_\\Documents/C2CCC\_TR\_2050\_Spectrum\_Needs.pdf}}, with modifications, enhancements, and corrections. 
It is not exhaustive; additional applications are currently being developed, and further ones will be developed in the future.

\begin{table*} 
    \caption{Overview C-ITS applications} \label{table_Chapter_ITSProtocols-TP-ITSAppOverview} 
    \centering
    \thispagestyle{empty}
    \begin{tabular}{p{3.5cm}p{13.5cm}}
	\toprule
	\textbf{Category} & \textbf{Applications} \\ 
	\midrule
	Traffic safety avoidance & Traffic jam ahead warning; Hazardous location warning; Emergency vehicle warning; Emergency brake light; Slow Vehicle Warning; Stationary vehicle warning; Overtaking warning; Intention sharing; Overtaking assistance; Overtaking assistance advances (including motorcycle); Collision risk warning; Intersection collision warning;  Wrong-way driving warning; Motorcycle approaching indication \\ \hline
	Cooperative Awareness & Behavior CAM; Road status (holes in the road, etc.) by infrastructure; Driver status CAM; Vehicle stats CAM; High-Definition sensor sharing; See-through for passing \\ \hline
	Vehicular automation & Basic ACC (level 2); Basic (level 2-3) C-ACC; Advanced (level 3-4) C-ACC (increase 20 Hz for CAM, CIM, ...); Basic (level 3-4) platooning (increase 20 Hz small CAM + Platoon Management); Advanced (level 4-5) platooning (including CIM + CPM camera/radar sensor data); Automated (level 4-5) vehicles (as advanced C-ACC + Camera/Radar sensor data); Basic merging assistant (inter-vehicular negotiations/roadside management); Advanced merging assistant (as Basic + increase $\Leftarrow$ 10 Hz small CAM); Automatic parking (basic and automated parking); Automation assist in tunnels (location precision assist); Automation level road assignment static and dynamic \\ \hline
	Road Works Warning & Short-term mobile; Basic short-term static (only road allocation awareness); Advanced short-term static (as basic + dynamic speed management depending on traffic density); Basic long-term static (only road allocation awareness); Advanced long-term static (as basic + dynamic speed management depending on traffic density); Emergency road works mobile (as short mobile with additional notifications) \\ \hline
	Traffic Flow & In-vehicle signage navigation (MAP-cloud services); In-vehicle signage local (dynamic or not managed by traffic management); Dynamic speed (direct + MAP-cloud service); Dynamic sign information (short-term direct + MAP-cloud service); Road topology (MAP) provisioning by authorities; Network flow optimization; Shock wave damping; Efficient traffic flow urban/highway; Complex lane marking; Regulatory/contextual speed limits notification; Traffic light optimal speed advisory; Zone access control for urban areas notification; Zone access control for urban areas enforcement; Enhanced route guidance and navigation; Public transport vehicle approaching; Greenlight optimal speed advice \\ \hline
	Intersection Safety & Red light violation warning; Energy-efficient intersection service; Stopping behavior optimization; Intersection obstacle indication; Queue warning; Left turn assist; Stop sign assist; Disabled vehicle warning \\ \hline
	Traffic Priority & Priority request business transport local; Priority request public transport local; Priority request emergency local; Priority request group of cyclists local; Priority request public transport via emergency center; Priority request emergency via emergency center; Priority request group of cyclists via the emergency center \\ \hline
	Vulnerable Road Users (VRU) & Bicycle safety awareness (CAM or CPM); Bicycle priority; Bicycle approaching indication; Pedestrian awareness (CAM or CPM); Motorcycle awareness (CAM) \\ \hline
	Traffic Information & Virtual variable message signs; Traffic information service; Virtual vehicle re-identification in traffic center \\ \hline
	Incident Management & Automatic incident detection (detection by vehicle); Automatic incident detection (detection by infrastructure); Incident Warning	 \\ \hline
	Navigation & Intermodal Route Planner; Standard Navigation; HD-MAP general map updates; HD-MAP local updates by vehicles and infrastructure for automated driving Strategic (cloud); HD-MAP local updates by vehicles and infrastructure for automated driving Tactile; HD-MAP and navigation map updates; Highway chauffeur (level 2-3); Rerouting; Eco route planner; Basic parking assist (directions); Advanced parking assist (specific parking lot); Geographical location improvement info exchange (PoTi) \\ \hline
	Media & Point of interest notification; ITS local electronic commerce; Media downloading; Multimodality support; Information on alternative fuel vehicles (AFV) fueling and charging stations \\ \hline
	Vehicle Service & EV charging point planner; Insurance and financial services; pay how you drive; Probe vehicle data; IMMA interface; Fleet management; Loading zone management; Automatic access control and automatic parking access; Road tolling; E-Tachograph\\ \hline
	Railway & Railway-road crossing; Urban rail safety \\ \hline
	Security and Privacy & Security key updates \\ \hline
	System Operations & Vehicle software/data provisioning and update; Vehicle and RSU data calibration and system management; Vehicle and RSU data calibration and system management ITS-G5 specific; ITS system management \\
 \bottomrule
 
\end{tabular}
\end{table*}

\subsubsection{ITS Application Support Functions}
\label{Chapter_ITSProtocols-TP-ITSAppSupport}

For any applications to function correctly and efficiently, certain services are required to facilitate the acquisition and distribution of information.
ITS defines the implementation of several communication-related support services.
These include data descriptions, service identification, data handling, and distribution mechanisms.
All applications and all facilities functionalities are specified independently at the lower layers and the communication technology used.
They operate on direct ad-hoc, cellular, broadband, or other communication technologies.

In essence, the facilities layer provides interfaces for the applications.
This encompasses a multitude of elements, including information from the traffic infrastructure (local and TMC) for RSU, on-board data for vehicles, communication profiles, subscription management for messages and information, the management interface for local station management and external management from, e.g., a TMC or vehicle manufacturer, information visualization, addressing, PoTi, and security.
In accordance with ETSI (\cite{TS102894-1}), the aforementioned requirements are sorted in the four groups management, application support, information support, and communication support.
This specification is linked with the preceding chapter \ref{Chapter_ITSProtocols-TP-ITSAppOverview}, which outlines the application requirements.
Additionally, the support functions are related to the Local Dynamic Map (LDM) \cite{EN302895} \cite{EN18750}, a database that holds information about the environment and messages received. 
It can be extended to function as an information exchange facility.
The LDM is sometimes described in different layers to distinguish between static information (such as maps), information that change at a low rate (such as weather and street signs), dynamic information (such as traffic light information, variable street signs, and traffic flow), and safety-relevant information (such as obstacles and driving maneuvers).
ISO facilities \cite{TS17429} focus on support functions for communication profiles, content subscriptions, and service handling.

The data elements and the description of the environment are essential for the understanding of ITS services.
ETSI specifies the data, the Common Data Dictionary (CDD), which should be used to define the message set, thus providing a reference for further development \cite{TS102894-2}.
ISO specifies a spatio-temporal data dictionary \cite{TR21718}, which includes specifications for static and dynamic data for all types of connection- and automation-related applications.
In SAE, the definition of the message set and the data elements are included in the same document \cite{SAEJ2735}.
Some descriptions include Position and Timing (PoTi) values.
Accurate timing and positioning are essential for all safety-related applications.
The improvement of location awareness and accurate positioning \cite{EN302890-2} in all dimensions (longitudinal, lateral, elevation, time) is fundamental for ITS applications.

Identification of services and stations at all layers is required for the reliable information exchange.
Different layers use different sets of identifiers.
On the application layer, the Application Object Identifier (AID) \cite{TS102965} or globally unique identifiers \cite{EN17419} are used to identify different applications.
New applications must be registered on the ISO web page\footnote{ISO TS 17419 application number assignment: \url{https://standards.iso.org/iso/ts/17419/TS17419\%20Assigned\%20Numbers/}}.
The website also provides a list of all currently assigned numbers. 
It includes numbers assigned by CEN (on behalf of ETSI), IEEE, and ISO.
However, the list was last updated in the year 2016.
At the transport layer, applications are identified by port numbers, such as those given in \cite{TS103248}.
Depending on the network layer used, different identification schemas can be used.
For example, GeoNetworking \cite{EN302636-1} can only transport BTP, IPv6, or any unspecified protocol according to \cite{EN302636-4-1}.
Identifications on lower layers are usually media-dependent and specify the network protocol used.

C-ITS is designed to work with different communication technologies and different channels \cite{TR103439}.
For this purpose, ETSI specified the communications architecture for Multi-channel Operation (MCO)\cite{TS103696} based on the station architecture \cite{EN302665}.
The functionality is divided into specifications for general requirements \cite{TS103697}, the facilities layer including DCC \cite{TS103141}, the networking layer \cite{TS103836-4-1}, and the access layer \cite{TS103695}.
All specifications are linked together for such a multichannel system.
The specifications are independent of the technologies or channels used.
An overview is presented is given in \cite{Bazzi2024}.

\subsubsection{ITS Application Message services}
\label{Chapter_ITSProtocols-TP-ITSAppMessages}

Several message protocols have been and will be defined based on the applications described in section \ref{Chapter_ITSProtocols-TP-ITSAppOverview}.
These specifications include the message service, which describes the use of the protocol and the conditions under which it should be used.
As mentioned earlier, the focus is on the messages used in Europe.
For comparison, US message definitions are mentioned where appropriate. 
Section \ref{Chapter_ITSProtocols-TP-ITSApp-PM} shows the correlation between these messages in a compact form.

The definition of the message is based on the specification in the corresponding documents.
For messages defined by ETSI and ISO and relevant to Europe, the ASN.1 description is publicly available on the ETSI server\footnote{ETSI forge git repository: \url{https://forge.etsi.org/rep/ITS/asn1}}.
For the message specification, it is important to understand that the specifications are divided between ETSI and ISO.
ETSI is responsible for applications based on traffic participants and ISO for applications based on infrastructure participation.
Both organizations define most of the message with different specifications. 
An exception is ETSI TS 103 301 \cite{TS103301}, which is closely related to ISO TS 19091 \cite{TS19091}.
This specification includes several messages related to infrastructure information provisioning.
The messages from the USA are included in the SAE J2735 \cite{SAEJ2735} specification.

The messages themselves are not the applications.
The applications use the message to communicate application-related information from one station to one or more other stations.
The specifications include the interface that applications can use to construct and send messages, management information, possible triggering conditions, etc.
To improve comparability, the C2C-CC and C-Roads consortia specify profiles called Basic System Profile (BSP) \cite{C2CCCBSP} and Harmonized Communication Profile \cite{CROADSHCP}, respectively.
These profiles specify the interpretation of the specifications and define the use of optional or negotiable parts.

\paragraph{Cooperative awareness message}
\label{Chapter_ITSProtocols-TP-ITSApp-CAM}
The Cooperative Awareness Message (CAM) is the basic service message defined in ETSI EN 302 637-2 \cite{EN302637-2}.
In an ITS it is essential to know that other stations are nearby.
This includes position and dynamic (e.g., speed, steering angle) and static attributes (e.g., vehicle dimensions, weight).
\cite{EN302637-2} focuses on the vehicle specification of the CAM, but introduces all the means for the CAMs for other stations. 
CAMs are sent out periodically depending on the speed and the environment.
All stations, especially vehicles, should send CAMs at all times. 
If no message is sent for some time, then the network layer sends out beacons \cite[p. 37]{EN302636-4-1} so that other stations can become aware of that station.
These beacons contain only the basic information, including the position vector.

\paragraph{Decentralized environmental notification message}
\label{Chapter_ITSProtocols-TP-ITSApp-DENM}
The Decentralized Environmental Notification Message (DENM) is specified in ETSI EN 302 637-3 \cite{EN302637-3} for \textit{Release 1} and ETSI TS 103 831 \cite{TS103831} for \textit{Release 2}.
DENMs distribute information about road hazards, unusual traffic, and unusual road conditions.
They are valid for a specific area of relevance specified in the message.
To distribute a DENM, a distribution area is specified and the message should be forwarded to the relevance area and within the transmission area.

The following three examples are provided to give insight into possible applications.
\begin{itemize}
	\item \textbf{Emergency Vehicle Alert}: When an emergency vehicle is on a call, it can be either en route to a call or at the call. In the former case, the vehicle needs to alert other vehicles of its presence and intentions so that the emergency vehicle's path can be cleared. This is especially important when crossing an intersection with oncoming traffic. In the latter case, the emergency vehicle and the personnel are at the scene. Other road users should be informed that an incident is in progress and that special care is required. Emergency vehicles can authenticate themselves with a special authorization included in their security certificate.
	\item \textbf{Intersection Collision}: There are two types of intersections with very different safety concerns: signalized and non-signalized. The non-signalized can be further divided into those with traffic signs and those  without traffic signs. In addition, intersections can be located either in urban or rural areas. Any vehicle, infrastructure, and other station can send an alert to detect a possible collision or a vehicle entering an intersection without right of way or a traffic light violation.
	\item \textbf{Road Side Alert}: Notification messages sent by the roadside can be considered as a separate category. This includes alerts forwarded by other road users. The majority of messages are likely to be infrastructure generated. Examples include warning of traffic jams, ice on the road, trains, or weather conditions. These messages are relevant when sent by a public authority and should therefore only be sent by the infrastructure with high confidence. On the other hand, other road users can rely on this information. 
\end{itemize}

\paragraph{Electric vehicle information message}
\label{Chapter_ITSProtocols-TP-ITSApp-EV}
Electric vehicles need to be charged.
Currently, these charging points are still rare. 
Therefore, in 2012, ETSI specified the Electric Vehicle Charging Spot Notification (EVCSN) \cite{TS101556-1}.
This specification allows the announcement of charging stations and their charging spots, including information about  opening hours, availability, possible waiting time, and bookability (called blocking).
The specification includes a mechanism to announce any Point of Interest (POI), but only electric vehicle charging points are specified.
A POI specification should have been made in an additional specification, and EVCSN should have used that specification.
In the current situation, \cite{TS101556-1} would need to be updated if a new POI is added.

The related specification is the 2014 Electric Vehicle Reservation (EV-RSR) \cite{TS101556-3}.
The protocol provides a mechanism to reserve a charging spot depending on the needs of the vehicle.
The protocol includes a feedback mechanism to confirm a reservation.
Unfortunately, there is no interaction between the notification and the reservation.
The same identification objects are not used for the charging station and the charging spot.

Both messages were not changed after the initial specification, and an update would be needed for real-world deployment.
In particular, the interaction between both messages and the payment part would need to be reworked.

\paragraph{Interference management zone message}
\label{Chapter_ITSProtocols-TP-ITSApp-IMZM}
The ITS frequency band (see section \ref{Chapter_ITSProtocols-TP-D-DSRC}) is adjacent to other bands, including DSRC for tolling.
The legacy systems do not follow the same strict bandwidth and spectrum requirements that are mandatory today. 
To protect such systems, mechanisms have been introduced by ETSI to make stations aware of such systems.
Two possibilities are standardized for this purpose: $\alpha$) using a map with all protection zones, $\beta$) the legacy system sends out protection information.
For the latter, the Interference Management Zone Message (IMZM) was standardized in ETSI TS 103 724 \cite{TS103724}.
This message contains information on protection zones.
For these zones, the frequency to be protected and the mitigation techniques are described depending on the technology used.
ITS participants in these zones shall follow the rules provided by the IMZM.

\paragraph{In-Vehicle information message}
\label{Chapter_ITSProtocols-TP-ITSApp-IVI}
Road traffic management is highly dependent on roadside signage. 
This information can be transmitted using the In-Vehicle Information Message (IVIM) specified in ETSI TS 103 301 \cite{TS103301}.
The specification uses the protocol definition from the 'dictionary of in-vehicle information (IVI) data structures' as specified in ISO TS 19321 \cite{TS19321}. 
ETSI TS 103 301 specifies only dissemination parameters on the different layers.
ISO TS 19321 itself uses several other specifications.
The coding of traffic signs is specified in the 'graphic data dictionary' in EN ISO 14823 \cite{EN14823} and its successor EN ISO 14823-1 \cite{EN14823-1}.
Vehicle identification is specified in three documents: $\alpha$) unique identification of road vehicles in ISO EN 24534-3 \cite{EN24534-3}, $\beta$) global unique identification in ISO TS 17419, now superseded by ISO EN 17419 \cite{EN17419}, and $\gamma$) numbering and data structures for automatic vehicle and equipment identification in ISO EN 14816 \cite{EN14816}.
Payment-related information for DSRC is specified in ISO EN 14906 \cite{EN14906}.
Finally, aspects related to signalized intersections are taken from ISO TS 19091 \cite{TS19091}.

\paragraph{Map data}
\label{Chapter_ITSProtocols-TP-ITSApp-MAP}
The information about road topologies in general and intersections in particular can be provided by the Map Data (MAP) defined in ETSI TS 103 301 \cite{TS103301} as Road and Lane Topology (RLT).
The message itself is called MAP (topology) extended message MAPEM.
ETSI TS 103 301 is just a wrapper for the ISO TS 19091 \cite{TS19091} specification. 
This is in turn a profile of the original specification in SAE J2735 \cite{SAEJ2735}.
Furthermore, values from \cite{EN24534-3} and \cite{EN14816} are employed.
With the topology data it is possible to describe lanes and traffic flow relationships at intersections and junctions.
This information can be used at signalized intersections to inform vehicles and adjust the timing based on this traffic.
The map data can be used to determine right of way and calculate possible traffic violations and accidents at intersections that are not signalized.

\paragraph{RTCM corrections message}
\label{Chapter_ITSProtocols-TP-ITSApp-RTCM}
The position of traffic participants in the context of ITS is mainly derived from GNSS.
Satellite navigation systems are very sensitive to weather conditions and obstacles blocking the line of sight to the satellite.
Correction signals from a base station can be used to counteract these distrubances.
These base stations have accurate position knowledge and use this data to calculate the derivation of the GNSS measurement to the actual position.
The derivations can be transmitted to (mobile) clients so that they can use the information to improve their position.
The Radio Technical Commission for Maritime Services Corrections Extended Message (RTCMEM) is specified in ETSI TS 103 301 \cite{TS103301} as GNSS Positioning Correction (GPC).
Only service-related information is part of the specification.
For the protocol itself, ETSI TS 103 301 is a wrapper for the ISO TS 19091 \cite{TS19091} specification. 
This is in turn a profile of the original specification in SAE J2735 \cite{SAEJ2735}.
Furthermore, values from \cite{EN24534-3} and \cite{EN14816} are employed.

\paragraph{Signal phase and timing message}
\label{Chapter_ITSProtocols-TP-ITSApp-SPaT}
At an intersection, traffic lights can provide information to approaching vehicles.
This information includes the signal plan parameters and the traffic light's switching cycle generated by the TLC.
The data elements are included in the Signal Phase and Timing Extended Message (SPATEM).
The data is linked to the MAP data (see \ref{Chapter_ITSProtocols-TP-ITSApp-MAP}) to provide information on the status of the traffic light for each lane .
Information on the expected subsequent cycles is also included.
It should be noted that these expected values are only solid for fixed-time programs of the TLC.
In most cases the programs are traffic-adaptive.
This means that the phases can glide within a predefined window and the TLC calculates the best timing based on the actual traffic conditions.
These calculations may change every second.
Prediction is currently part of the product development of many traffic light manufacturers and suppliers.
Based on the SPATEM, vehicles can calculate the optimal speed to pass through the intersection without stopping.
This reduces noise and $CO_2$ emissions and improves traffic flow.
For the protocol itself, ETSI TS 103 301 is just a wrapper for the ISO TS 19091 \cite{TS19091} specification. 
This is in turn a profile of the original specification in SAE J2735 \cite{SAEJ2735}.
Furthermore, values from \cite{EN24534-3} and \cite{EN14816} are employed.

\paragraph{Signal request message}
\label{Chapter_ITSProtocols-TP-ITSApp-SRM}
Special vehicles may request priority at an intersection.
These vehicles may include public transportation vehicles or emergency vehicles.
Both should be able to pass an intersection faster than normal vehicles, but at the same time, safety should always be ensured.
With the Signal Request Extended Message (SREM), these vehicles can request a green light to pass an intersection.
The TLC will incorporate those data and attempt to optimize the signal planning accordingly.
It is not permitted to make ad-hoc changes to signals due to safety regulations and some minimum time restrictions (e.g., \cite{Ril15})
With regards to the protocol itself, ETSI TS 103 301 is merely a wrapper for the ISO TS 19091 \cite{TS19091} specification. 
This is in turn a profile of the original specification in SAE J2735 \cite{SAEJ2735}.
Furthermore, values from \cite{EN24534-3} and \cite{EN14816} are employed.

\paragraph{Signal status message}
\label{Chapter_ITSProtocols-TP-ITSApp-SSM}
The Signal Status Extended Message (SSEM) enbales the RSU to convey feedback regarding the reception of the SREM.
This information can then be utilized by the vehicle in question to adjust its driving behavior accordingly. 
In the case of public transportation vehicles, this may necessitate a stop, while emergency vehicles may require a more cautious approach when traversing the intersection. 
With regards to the protocol itself, ETSI TS 103 301 is merely a wrapper for the ISO TS 19091 \cite{TS19091} specification. 
This is in turn a profile of the original specification in SAE J2735 \cite{SAEJ2735}.
Furthermore, values from \cite{EN24534-3} and \cite{EN14816} are employed.

\paragraph{Service announcement message}
\label{Chapter_ITSProtocols-TP-ITSApp-SA}
The ITS architecture can facilitate the integration of diverse communication technologies (e.g., \cite{IEEE1609-4}) and different channels within the same technology (MCO). 
To address the distribution of different services across different technologies and channels, the Service Announcement Service (SAS) is specified in \cite{TS102890-1} and \cite{EN302890-1}. 
The SAS is based on ISO TS 16460 \cite{TS16460} and EN ISO 22418 \cite{EN22418}. 
The latter is aligned with ETSI EN 302 890-1 \cite{EN302890-1}.

The SAS allows a service provider to announce services it can offer for other stations (referred to as a service announcer). 
It defines the interfaces for the transmission and reception of Service Announcement Messages (SAM) and the security service. 
The SAS can be used locally on the same station, separately, or in combination.
It is recommended that service providers register with the management entity of the ITS reference architecture (as defined in \cite{EN302665}) to register, update, or deregister their ITS service.
Similarly, it is recommended that service users also register, update, or deregister for services of interest. 
The management entity may notify registered service users about the reception of a SAM.

Finally, it is recommended that the distribution of services be based on the geographical area in which the service is relevant. 
This is not limited to the relevance area but extends to a dissemination area, where a station can consume services relevant to its current position and those areas it plans to traverse during its journey. 
The services that can be announced are safety-related or non-safety-related. 
They include information about the technology (only IEEE-802.11-based technologies are supported currently). 
From a protocol perspective ETSI EN 302 890-1 \cite{EN302890-1} supports only WAVE Short Message Protocol (WSMP), Geonetworking (GN), Basic Transport Protocol (BTP), and IPv6 based on TCP in accordance with IEEE 1609.3 \cite{IEEE1609-3}. 

The dissemination area (in compliance with ETSI EN 302 931 \cite{EN302931}) for a SAM is specified by the ITS service this is announcing the service. 
ITS services with different dissemination areas should be placed in different SAMs. 
In the event that BTP is utilized, the BTP port number should be as specified in ETSI TS 103 248 \cite{TS103248}. 
The SAM should be sent as a broadcast with a repetition interval between 0 and 255 (no unit is provided). 
The SAM may be transmitted via ITS-G5 in accordance with communication parameter setting $CPS_001$ from ETSI TS 103 301 \cite{TS103301}.
The ITS-AID value of the SAS should be as specified in ETSI TS 102 965 \cite{TS102965} and the SAS should utilize the SSPs as specified in IEEE 1609.3 \cite{IEEE1609-3}. 
While data confidentiality is not a security concern, data integrity, non-repudiation, and source authentication are required. 
According to ETSI TS 103 097 \cite{TS103097}, a generic profile should be employed.

To differentiate between various service classes, a priority framework is essential. 
In \cite{SERAN2016}, four service classes are proposed: safety-critical, legislation, security, and non-security. 
The safety channel G5A should not be utilized as a reference for a service, but it should be employed to disseminate the SAM messages. 
The designation SAM was subsequently altered to Service Announcement Extended Message (SAEM) to distinguish the European version from the US version of the specification.

\paragraph{Tire information system message}
\label{Chapter_ITSProtocols-TP-ITSApp-tire}
The safety of a vehicle is contingent upon the proper inflation of its tires.
ETSI has defined the Tire Information System (TIS) and the Tire Pressure Gauge (TPG) for this purpose in TS 101 556-2 \cite{TS101556-2} as instrument for measuring, monitoring the tire pressure, respectively.
TIS facilitates the exchange of pressure measurement information between the tire, vehicle system, and driver.
The TPG is responsible for exchanging that information with other vehicles or services provided. 
The TPG will facilitate the automatic refilling process.

\paragraph{VRU awareness message}
\label{Chapter_ITSProtocols-TP-ITSApp-VAM}
Vulnerable Road Users (VRUs) such as pedestrians, cyclists, motorcyclists, wheelchairs, or animals require special protection.
These uses do not send CAMs and are therefore not communication visible.
The Vulnerable Road Users Awareness Message (VAM) is defined in ETSI TS 103 300-2 \cite{TS103300-2} and TS 103 300-3 \cite{TS103300-3}.
A particular type of VRU is the motorcyclist.
They send CAMs and should not send VAMs.
VRU awareness messages should only be sent by stations that are VRU.
As VRU may occur in groups, and the communication channel could be overloaded with every VRU sending VAMs individually, the VAM includes a mechanism to cluster VRU together. 
Additionally, clusters elect cluster leaders.
Only those leaders send VAM, including the cluster information.
Clusters are built and changed dynamically depending on the environment and number of VRU in the vicinity.

\paragraph{Platooning control message}
\label{Chapter_ITSProtocols-TP-ITSApp-PCM}
Platoons of vehicles, particularly trunks, can be formed when driving through urban areas or on motorways.
Platoons can be established if the platoon members have the same destination or share the same route for a perode of time.
The platoon then assumes control of the vehicles, with all members performing the same action (acceleration, deceleration, steering) in a coordinated manner.
The formation of platoons has the potential to reduce fuel consumption due to the possibility of reducing the gap between vehicles below the current safety standards.
However, in order for platoons to be efficient, regulations must be amended to permit such behavior.
Platoons will only be viable with automated vehicles due to the reaction times that would be required for human intervention given the small gap between vehicles.
The Platooning Control Message (PCM) idea is defined by ETSI in TR 103 298 \cite{TR103298}.
The specification process was halted in 2022 due to the necessity for further research to provide a foundation for standardization.
The outcomes of the European project ENabling SafE Multi-Brand pLatooning for Europe (ENSEMBLE)\footnote{ENSEMBLE in the CORDIS database: \url{https://cordis.europa.eu/project/id/769115/de}} were intended to substantiate this foundation, yet the endeavor was unsuccessful. 
PCM, in essence, can be utilized to create, join, leave, and maintain a platoon.

\paragraph{Collective perception message}
\label{Chapter_ITSProtocols-TP-ITSApp-CPM}
In a connected vehicular environment, exchanging information about objects or other traffic participants is possible.
Vehicles and infrastructure systems are equipped with a multitude of sensors, including cameras, radars, and lidars. 
These sensors are employed for surveillance of the environment and detection of other road users and objects.
The data collected can be shared using the Collective Perception Message (CPM) specified by ETSI in TS 103 324 \cite{TS103324} and motivated by TR 103 562 \cite{TR103562}.

\paragraph{Maneuver coordination message}
\label{Chapter_ITSProtocols-TP-ITSApp-MCM}
The automation of vehicles is introduced in a gradual manner.
For automated vehicles, some form of coordination between vehicles is necessary.
This is relevant in a number of traffic conditions. 
For instance, a vehicle may wish to drive on a motorway, or a vehicle may be required to turn left, in that cases it must coordinate with the traffic on the road.
Another example would be a system such as the Traffic Collision Avoidance System (TCAS) for planes. 
The system is designed to alert planes about other nearby airplanes and reduce the risk of mid-air collisions.
It can automatically provide individual flying maneuvers for each plane involved.
Such a system would be implementable with two vehicles on a possible collision course.
The process follows the four steps: $\alpha$) status sharing, $\beta$) intent sharing, $\gamma$) agreement seeking, and $\delta$) prescription.
ETSI defines the Maneuver Coordination Message (MCM) based on TR 103 578 \cite{TR103578} and TS 103 561 \cite{TS103561}.
The use cases are based on the definitions and the taxonomy defined in SAE J3216 \cite{SAEJ3216}.

\paragraph{Multimedia content dissemination}
\label{Chapter_ITSProtocols-TP-ITSApp-MCD}
An additional feature is specified in ETSI TS 103 152 \cite{TS103152}, the dissemination of multimedia information (MCD).
Multimedia data (video, audio, pictures, URL, etc.) can be used in both safety and non-safety application scenarios.
In safety cases, multimedia data can assist the driver or the automated driving system in assessing the environment.
In non-safety cases, multimedia information can be used for entertainment purposes.
The multimedia information can be related to CPM or DENM.
It can be transmitted multicasted via GeoNetworking to several peers.
Additionally, a peer-to-peer solution is foreseen, but only via GN6.

\paragraph{Diagnosis, logging, and status message}
\label{Chapter_ITSProtocols-TP-ITSApp-DLS}
The introduction of Rail2X to standardization is described in ETSI TR 103 694 \cite{TR103694}, which outlines use cases regarding maintenance and remote diagnosis of infrastructure systems.
The idea is that trains collect data along the track.
Subsequently, the information is transmitted to a responsible station or train yard server.
The corresponding message would is the Diagnosis, Logging, and Status Message (DLSM).
Standardization is ongoing with other use cases, such as trains introducing themselves at crossings to vehicles (Rail2Vehicle) or trains communicating with each other to exchange information, e.g., for coupling or decoupling (Rail2Rail).
Conversely, the protocol could be employed in the vehicular environment to facilitate the provision of wireless maintenance information to garages.

\subsubsection{Protocol matrix}
\label{Chapter_ITSProtocols-TP-ITSApp-PM}
The definition of messages in Europe is primarily conducted independently for V2V communication, though it is often based on definitions established in North America for the V2I communication.
Consequently, the standardization institutions ETSI, ISO, and SAE have developed a shared approach to message definition. 
In the United States, IEEE is responsible for specifying access, network, and transport-related topics, with the except of cellular radio.
Conversely, the SAE is responsible for defining all application-related topics and requirements.
SAE J2735 \cite{SAEJ2735} comprises all messages defined in the United States in one specification.

Table \ref{tab_protocol-matrix} displays the relationship between messages and the different message names.
The mapping is not always congruent.
If this is the case, it is indicated by the word 'partly.'
Messages that were not introduced before are explained afterward.
The order of the messages is derived from the order in SAE J2735 \cite{SAEJ2735}.

\begin{table*}
	\centering
	\caption{Protocol matrix}
	\label{tab_protocol-matrix}
	\begin{tabular}{p{5cm}p{8 cm}}
        \toprule
		\textbf{SAE}                     	  & \textbf{ETSI and ISO}\\ 
        \midrule
		Basic Safety Message BSM	   	       & Cooperative Awareness Message (CAM) \\
		Common Safety Request (CSR)            & no equivalent \\
        Emergency Vehicle Alert (EVA)          & Decentralized Environmental Notification Message (DENM), partly\\
        Intersection Collision Avoidance (ICA) & Decentralized Environmental Notification Message (DENM), partly\\
        Map Data (MAP)                    	   & MAPEM\\
        NMEA corrections (NMEA)				   & no equivalent \\
        PersonalSafetyMessage (PSM)			   & VRU Awareness Message (VAM) \\
        Probe Data Management (PDM) 		   & no equivalent \\
    	Probe Vehicle Data (PVD)			   & no equivalent \\
		Road Side Alert (RSA)				   & Decentralized Environmental Notification Message (DENM)\\
		RTCM corrections (RTCM)                & RTCM corrections extended message (RTCMEM) \\
		Signal Phase And Timing Message (SPAT) & Signal Phase And Timing Extended Message (SPATEM) \\
		Signal Request Message (SRM)		   & Signal Request Extended Message (SREM) \\
		Signal Status Message (SSM)			   & Signal Status Extended Message (SSEM) \\
		Traveler Information Message (TIM)     & In-Vehicle information message (IVIM) \\
		Test Messages          				   & no equivalent \\
		no equivalent          				   & Electric Vehicle Charging Spot Notification (EVCSN) \\
		no equivalent          				   & Electric Vehicle Reservation (EV-RSR) \\
        no equivalent          				   & Interference Management Zone Message (IMZM) \\
		no equivalent          				   & Service Announcement Message (SAM) \\
		no equivalent          				   & Tire Information System (TIS) \\
		no equivalent          				   & Collective Perception Message (CPM) \\
		no equivalent          				   & Maneuver Coordination Message (MCM) \\
		no equivalent          				   & Multimedia Content Dissemination (MCD) \\
		no equivalent          				   & Diagnosis, Logging, and Status Message (DLSM) \\
		\bottomrule
	\end{tabular}
\end{table*}

The Basic Safety Message (BSM) is equivalent to the European CAM.
It is a basic awareness service that provides information (position, identity, etc.) about the sending station at a frequency of up to $10 Hz$.

The Common Safety Request (CSR) essage allows stations sending BSM to request more information from other stations.
This request is sent as a unicast from a station that needs more information to realize a safety application.
The answer is provided in an extended version of the BSM and sent as a regular broadcast.

Emergency vehicles utilize Emergency Vehicle Alert (EVA) messages to inform other traffic participants of their imminent arrival or current location in the vicinity of an incident.
These message includes a Roadside Alert Message (RVA) and extend it with information about the emergency vehicle and the nature of the emergency.

In the event of a critical situation at an intersection, an Intersection Collision Avoidance (ICA) message can be sent from any station to inform all other stations about the situation.
This message includes information about the vehicles involved and the critical situation.

The Probe Data Management (PDM) message is utilized by a RSU.
It is transmitted to mobile stations to inform them of the vehicle data they are to collect and transmit to the RSU.
The objective of the collection is to generate communication and a vehicular traffic overview. 
The initial collection, for instance, can be employed to determine the communication range of an RSU.

Probe Vehicle Data (PVD) messages are transmitted primarily through mobile stations and include information about the path history of the station and other related values of this station.
These data are collected in snapshots, each including some part of the history of the path.
They are sent in response to the PDM message but can be sent without the probing.

The Road Side Alert (RSA) message is the equivalent of ETSI's DENM.
It holds information about nearby hazards for traffic participants.
The information is encoded in a manner that is as straightforward as possible, with the objective of minimizing the requirements for understanding the message and maximizing the number of stations that can understand the message.

Test messages can be used for the development of new services. 
Currently, SAE defines 16 messages that can be used for any purpose.
It is recommended that common framework and data types already defined be reused.

As a special message, the service announcement (see section \ref{Chapter_ITSProtocols-TP-ITSApp-SA}) message is not defined by SAE but rather by IEEE.
The WSMP Service Announcement \cite{IEEE1609-4} is equivalent to the European Service Announcement Message (SAM).

\subsection{ITS security and privacy}
\label{Chapter_ITSProtocols-TP-ITSApp-secp}

The ITS protocols and standards are designed with the safety of road users as a primary consideration.
Security and privacy are of paramount importance in the development of ITS.
Without security, there can be no safety. 
The ITS communication architecture standard design principles include, for example, user needs, communication capacity, reliability, availability, hybrid communication, flexibility, and \textit{communication privacy} \cite[p. 14]{EN302665}. 
Security is not mentioned as a design principle.
However, a comprehensive chapter is dedicated to security, which is integrated into the initial reference architecture.

The GeoNetworking standard in section 9.1.7 \cite[p. 21]{EN302636-3}, asserts that secure communication based on digital signatures and certificates is mandatory.
The stated objectives are to provide means for authentication, authorization, integrity, non-repudiation, plausibility checks, rate limitations, and trustworthiness assessments. 
Furthermore, measures for the protection of privacy are mandatory in order to protect the user. 
Therefore, pseudonyms and anonymous certificates should be employed. 

The European General Data Protection Regulation (GDPR) represents one of the fundamental rights in the European Union. 
Every communication technology and application must be evaluated against the GDPR requirements. 
In 2016, the EC conducted a study\footnote{\textit{C-ITS platform} final report published: \url{https://transport.ec.europa.eu/transport-themes/intelligent-transport-systems/cooperative-connected-and-automated-mobility-ccam_en}} regarding privacy. 
In this study, only CAM and DENM were identified, and the study focused on a single communication channel. 
However, it is also necessary to consider MCO and hybrid communication, as well as the implementation of a data protection impact assessment. 
In 2017, the \textit{C-ITS platform Phase II} report\footnote{Full report on the EC website: \url{https://transport.ec.europa.eu/document/download/35a339a8-03e4-484f-9e63-1f14a192bb3c_en?filename=2017-09-c-its-platform-final-report.pdf}, accessed 2024-06-13} identified tracking as a risk and certificate change as a mitigation technique. 
The document emphasizes the necessity of developing and deploying measures to minimize user tracking.
Therefore, an in-depth analysis is required to identify potential possible risk factors and the development of mitigation techniques such as do-not-track and encryption is essential.

As outlined in \cite{ikopaD3.3}, five key risk factors pertaining to privacy must be taken into account:
\begin{itemize}
	\item \textbf{Identifiers must be changeable:} The alternation of identifiers must be feasible on any communication channel and technology prior to commencing a new transmission.
	\item \textbf{All identifiers of a single stack need to be changed in coordination:} The linking of identifiers can only be prevented when all changes occur simultaneously on all layers. This includes MAC addresses, session identifiers, IP addresses, pseudonyms, and so forth.
	\item \textbf{Multi-stack applications need to coordinate identifier changes across stacks:} In the event that communication is related to disparate technologies or occurs via disparate channels for a given technology, it is imperative that all identifiers for related layers and technologies undergo simulatious alteration. 
	\item \textbf{Quasi-identifiers in the data content may enable linkability:} Pseudo-identifiers are identifiers that are indirectly related to the ITS station (ITS-S). Examples of pseudo-identifiers include a TCP port of a connection or the time and coordinates of a message. 
	\item \textbf{Identity beacons:} Communication technologies such as Bluetooth or WLAN periodically transmit information about their own status to announce their presence. This is also true for ITS-S, which transmit beacons if no other message is sent for some time. It is currently unclear if this is necessary for all channels.
\end{itemize}

ETSI has established a set of specifications regarding security and privacy.
The following list contains the most important documents within this set.

\begin{itemize}
	\item \textbf{TR 102 893:} A Threat, Vulnerability, and Risk Analysis (TVRA) was developed for V2V and V2I ad-hoc communication in the 5.9 GHz ITS band. The results and recommended actions are described in \cite{TR102893} in the context of the BSA.
	\item \textbf{TR 103 415:} The pre-standardization study on pseudonym change management \cite{TR103415} provides an overview of pseudonym change strategies and defines relevant parameters. An evaluation of different change strategies is scheduled to be added at a later point in time.
	\item \textbf{TS 102 731:} The building blocks for the ITS security services and the architecture for the pre-phase are defined in \cite{TS102731}. The final security architecture is specified in TS 102 940 \cite{TS102940}. TS 102 731 includes mechanisms for secure and privacy-protecting communication, such as authentication, authorization, and identity management.
	\item \textbf{TS 102 867:} The specification maps at stage 3 the security architecture and communication aspects defined in \cite{TS102731} against the IEEE 1609.2 \cite{IEEE1609-2} security features. It identifies gaps and describes the mitigation-related requirements \cite{TS102867}. Stage 3 is based on the three-stage model presented in ETS 300 387 \cite{ETS300387}, which is based on ITU Recommendation I.130 \cite{ITUT_I.130}. In brief, the following definitions apply: Stage 1 is the user view of the system. Stage 2 describes the system components. Stage 3 represents the protocols and interfaces.
	\item \textbf{TS 102 940:} The specification is based on the pre-work in TS 102 731 \cite{TS102731}. TS 102 940 \cite{TS102940} defines the ITS communications security architecture and entities, as well as the corresponding security management, which includes, but is not limited to, the PKI process, trust policies, and certificate management. Additionally, it specifies the security elements and services described in the ETSI system architecture \cite{EN302665}.
	\item \textbf{TS 102 941:} The trust and privacy management \cite{TS102941} specification identifies and specifies in detail the system aspects in regards to the trust between ITS stations and the necessary parts for the certificate and identity management.
	\item \textbf{TS 102 942:} The access to ITS-related services is limited to authorized users to ensure high quality and highly reliable system operation. The access control specification \cite{TS102942} delineated the associated services and measurement for secure and privacy-concerned communication.
	\item \textbf{TS 102 943:} The services for securing the communication and the insurance of confidential communication are specified in \cite{TS102943}. The application determines the level of confidentiality. Most of those do not require encrypted message delivery; rather, they rely on authorized message delivery.
	\item \textbf{TS 103 097:} The security headers, certificate formats, and corresponding data structures are specified in \cite{TS103097}. They are used to secure the communication between ITS stations. The specification is described as a profile of the IEEE 1609.2 security architecture \cite{IEEE1609-2}.
	\item \textbf{TS 103 600:} All security measures defined in \cite{TS102940}, \cite{TS102941}, and \cite{TS103097} have implications on the communication between ITS stations. Consequently, an interoperability test specification \cite{TS103600} has been defined to ensure the overall system functionality.
	\item \textbf{ISO TS 21177:} Furthermore, in addition to the ETSI specification, ISO specifies in TS 21177 \cite{TS21177} services for mutual authentication and data integrity between two trusted stations and within a station. These services work to fulfill the application, management, and safety requirements.
\end{itemize}

In contrast to the numerous specifications employed in Europe, the United States specifies the security-related aspects in the IEEE 1609.2 \cite{IEEE1609-2} specification. 
While this has the advantage of consolidating all relevant information in a single location, it is less flexible in terms of extensions, change, and reuse. 
It is also noteworthy that security in ETSI is addressed at all layers except the access layer. 
This approach distributes the security services over the ITS station with specific entities.
Conversely, the IEEE approach only addresses data at the application layer.

\section{ITS-related communication protocols} 
\label{Chapter_ITSProtocols-CommunicationProtocols}

The exchange of information between involved systems is a fundamental aspect of distributed applications.
The communication process is dependent on the specific needs of the applications in question and should enable them to fulfil their intended purpose.
Given the diverse range of applications, the use of a single communication technology is not a viable option.
Safety-related and time-critical applications, such as collision warnings or VRU protection, have distinct communication requirements compared to general applications, such as weather forecasts or map updates. 
In particular, when considering the functional safety perspective \cite{ISO26262-9}, the use of redundant and resilient communication is essential.

In the field of ITS, a multitude of communication specifications are employed.
This section illuminates some of these specifications and introduces the communication protocols utilized.

The ITS architecture is delineated in ETSI EN 320 665 \cite{EN302665}. 
It encompasses a comprehensive overview of the ITS landscape, a communication model, and a station reference model of four station types: central, vehicle, personal, and roadside. 
A detailed architectural description is provided in section \ref{Chapter_ITSProtocols-CP-ITSCom}. 
\cite{TR102962} provides an overview of the integration of cellular communication in the ITS architecture.

The security architecture and services, including standards of trust and privacy, access control, confidentiality, headers, formats, and TVRA, are described in detail in section \ref{Chapter_ITSProtocols-TP-ITSApp-secp}.

The application (see section \ref{Chapter_ITSProtocols-TP-ITSAppOverview}) and facility (see section \ref{Chapter_ITSProtocols-TP-ITSAppSupport}) parts are divided into the principle application description (\textit{application}) and the message description and support functions (\textit{facilities}). 
Both are mentioned for ISO and ETSI, but the content is often similar and identical.

The transport and network layer technologies and protocols are described in section \ref{Chapter_ITSProtocols-TP-Net}. 
Internet-based technologies like IPv6, UDP, TCP, MQTT, or AMQP are birefly introduced in section \ref{Chapter_ITSProtocols-CP-IP}.

The access layer and its technologies are divided into two parts: direct communication with DSRC (see section \ref{Chapter_ITSProtocols-TP-D-DSRC}) and C-V2X (see section \ref{Chapter_ITSProtocols-TP-D-CV2X}), and infrastructure-based protocols like cellular (see section \ref{Chapter_ITSProtocols-TP-I-Cel}). Additionally, there are other non-infrastructure-based protocols (see section \ref{Chapter_ITSProtocols-TP-D-Oth}).

\subsection{Internet-based protocols}
\label{Chapter_ITSProtocols-CP-IP}
The communication in ITS is, besides direct communication, based on the same principles as any other communication nowadays.
The following paragraphs present the most significant protocols and concentrate on their utilization in the field of ITS.
In addition to the fundamental protocols IP, TCP, and UDP, a selection of contemporary middleware protocols are also introduced.
The protocols described in the following sections (AMQP, CoAP, MQTT, SOAP, XMPP, and WebSockets) are employed in ITS for back-end and infrastructure communication.

\subsubsection[IP]{Internet Protocol (IP)}
\label{Chapter_ITSProtocols-CP-IP-IP}
The Internet Protocol (IP) is employed in a multitude of applications and services within the field of ITS.
For instance, communication within the stations is frequently conducted via IP.
Additionally, communication via cellular networks and the backend is based on the Internet protocol.
Currently, the sixth version (IPv6) \cite{RFC8200} is being deployed. 
The specification was released in 1998, and an increasing number of systems are now ready for use.
New protocols and developments in Europe are expected to be IPv6-capable\footnote{An overview about the deployment of IPv6 can be found on the portal of the European Commission: \url{https://ec.europa.eu/internet-standards/ipv6.html}}.
The GeoNetworking (see section \ref{Chapter_ITSProtocols-TP-Net-GN}) supports IPv6 only.

\subsubsection[UDP]{User Datagram Protocol (UDP)}
\label{Chapter_ITSProtocols-CP-IP-UDP}
The majority of applications based on ad-hoc communication that employ IP will utilize the User Datagram Protocol (UDP) \cite{RFC768} as their transport layer protocol.
This is due to the fact, in such an environment, information distribution cannot be expected to be mutual in many cases due to the rapidly changing physical conditions in traffic. 
Additionally, the majority of information is intended for one or multiple recipients. 
A lightweight transport protocol such as UDP is optimal for such IP-based information distribution.
Information distributed to numerous participants is, in the majority of cases, not subject to encryption requirements.
Mechanisms should be in place (e.g., message signatures, digital hashes, etc.) for the recipient to verify the sender's authenticity and the message content.
Additionally, messages are transmitted multiple times to address lost messages and transmission errors.
This can be accomplished at the application or facilities layer via application logic or at lower layers with ARQ or HARQ (Hybrid Automatic Repeat Request is ARQ in combination with FEC).
Such techniques are employed, for example, by 4th and 5th generation cellular networks or WLAN.

\subsubsection[TCP]{Transmission Control Protocol (TCP)}
\label{Chapter_ITSProtocols-CP-IP-TCP}
The Transmission Control Protocol (TCP) \cite{RFC793} is suitable for use in situations where reliable transmission is essential.
In the context of ITS, this is true for situations where one-to-one communication is needed.
Examples of such situations include the request and reception of new certificates, the transmission of personal information, or the usage of a specific service (e.g., charging point reservation).
In some cases, such communication additionally requires encryption and authenticity. 
TCP employs, among other techniques, ARQ for the reliable delivery of information.

\subsubsection[AMQP]{Advanced Message Queuing Protocol (AMQP)}
\label{Chapter_ITSProtocols-CP-IP-AMQP}
The Advanced Message Queuing Protocol (AMQP) is specified by OASIS \cite{AMQP1} and in an older version by ISO \cite{ISO19464}.
It is a byte-based application layer network protocol.
It provides different communication patterns, such as public-subscriber or point-to-point.
It uses a broker for the distribution of information.
AMQP is capable of transporting data from any application and environment.
In order to ensure reliability, it requires the use of a robust transport protocol, such as TCP.
Furthermore, it is able to leverage security mechanisms, such as Transport Layer Security (TLS), to enhance the security of the communication.

\subsubsection[CoAP]{Constrained Application Protocol (CoAP)}
\label{Chapter_ITSProtocols-CP-IP-COAP}
The Constrained Application Protocol (CoAP) \cite{RFC7252} is an IETF protocol for the web-based transport of information between constrained systems.
It is specified in accordance with the principles of HTTP and can be readily integrated with it. 
CoAP employs request/response mechanisms for data exchange based on REST principles.
It incorporates a service discovery concept and utilizes Unified Resource Identifiers (URI). 
It employs UDP and IPv6 over Low-Power Wireless Personal Area Networks (6LoWPAN).
6LoWPAN is an IETF protocol for Internt of Things (IoT) of CSMA/CA-based IEEE 802.15.4 networks (e.g., Zigbee).
The new home automation standard Matter\footnote{Connectivity Standards Alliance (CSA): Matter: \url{https://csa-iot.org/all-solutions/matter/}} utilizes DTLS, CoAP, and 6LoWPAN.

\subsubsection[XMPP]{Extensible Messaging and Presence Protocol (XMPP)}
\label{Chapter_ITSProtocols-CP-IP-XMPP}
The Extensible Messaging and Presence Protocol (XMPP) \cite{RFC6120} is a specification for the exchange of information and availability status based on XML streams.
It works on top of IP and TCP and provides a nearly real-time message delivery.
XMPP, formally known as Jabber, is commonly used for instant messaging applications but can be used to transfer any data.
In its most basic form, XMPP is a client-server architecture that can be extended to a peer-to-peer structure.
It functions on the basis of the request-response principle.
It provides, among other things, setup and teardown procedures, encryption, authentication, and error handling.

\subsubsection{MQTT}
\label{Chapter_ITSProtocols-CP-IP-MQTT}
MQTT (formerly Message Queuing Telemetry Transport) is a byte-based application layer network protocol for the of transport telemetry data for the IoT.
The protocol is specified by OASIS \cite{MQTT5} and, currently, in an older version with ISO \cite{ISO20922}.
Its design is intended to facilitate the transport of information from and to systems with limited computing power and low-bandwidth connections.
It is based on the publish-subscribe pattern.
The message broker is the central server that receives all messages and distributes them to the subscribed clients.
MQTT necessitates a dependable mode of information transport, which is why it frequently employs TCP and IP.
Nevertheless, any other reliable transport protocol or technology, such as Zigbee, can be utilized. 

\subsubsection{SOAP}
\label{Chapter_ITSProtocols-CP-IP-SOAP}
SOAP (formerly Simple Object Access Protocol) is an XML-based application layer messaging protocol for the conveyance of web service-related information.
SOAP is currently specified by the World Wide Web Consortium (W3C) in version 1.2 \cite{SOAP}.
SOAP can be used to transport data between distributed systems and to activate remote procedures.
It may use different communication protocols, such as HTTP, \textit{Simple Mail Transfer Protocol} (SMTP), TCP, or UDP.
It includes security features for encryption and authentication.
It can be extended by design.
SOAP defines the protocol, message, data, and communication concepts. 
Despite SOAP utilizing XML, it is not limited to this format.
In addition to XML, other formats such as BASE64, JavaScript Object Notation (JSON), and Comma-Separated Values (CSV) may be employed for data.

\subsubsection{WebSocket}
\label{Chapter_ITSProtocols-CP-IP-WebSocket}
WebSockets \cite{RFC6455} are defined application layer communication protocols.
They provide two-way communication for the exchange of information between a web browser and a web server.
WebSockets are not the same as HTTP.
However, WebSockets work over the well-known port 80 (unsecured) and 443 (secured) utilizing TCP.
They are compatible with existing HTTP infrastructure.
WebSockets may utilize the HTTP internal upgrade mechanism to change from ordinary HTTP to WebSockets.
With the help of WebSockets, a persistent connection is established, and client and server can send information at any time.
This represents an improvement to HTTP, where such a kind of exchange is not foreseen and can only be accomplished with the help of some stunts.

\subsection{ITS Communication protocols}
\label{Chapter_ITSProtocols-CP-ITSCom}

In ITS, several specialized communication-related protocols are described.
To understand the basic principles, architecture documents are created.
ETSI specifies its communication architecture, which is more a station architecture in \cite{EN302665}.
This specification is based on the ISO Communications access for land mobiles (CALM) specifications, which were created in the European research project CVIS\footnote{Co-operative Vehicle-Infrastructure Systems (CVIS): \url{https://cordis.europa.eu/project/id/027293/de}}.
ISO specifies in \cite{EN17427-1} the roles and responsibilities for C-ITS\footnote{The roles were developed based on the results of the German research project COmmunication Network VEhicle Road Global Extension (CONVERGE): \url{https://fgvt.htwsaar.de/site/en/converge/}}.

The communication in ITS can be direct or infrastructure-based.
Direct communication is described in section \ref{Chapter_ITSProtocols-TP-D}, while infrastructure-based communication is discussed in section \ref{Chapter_ITSProtocols-TP-I}.
The majority of ITS specifications are developed independently from the protocols of the underlying layers.
This approach is employed to facilitate the interchangeability between lower-layer protocols, despite their inherent differences.

The distinction between infrastructure and direct communication is not always clear-cut.
Technologies such as Bluetooth or RFID could be considered both, depending on the circumstances.
The exploitation of information from a Bluetooth device can be accomplished passively or actively. 
In the passive mode, only the existence of a Bluetooth device is detected. 
In the active mode, a connection is established.

\subsubsection{Network protocols}
\label{Chapter_ITSProtocols-TP-Net}

ETSI specifies technology-independent protocols on the transport and network layer, which is referred to in ETSI terminology as GeoNetworking and Transport layer. 
The IEEE WAVE protocol is a similar technology-independent protocol, although it is not entirely independent because it incorporates IEEE 802.11 as an access technology.
The following sections describe the network and transport layer parts of ETSI and IEEE.

\subsubsection[BTP]{Transport Protocol (BTP)}
\label{Chapter_ITSProtocols-TP-Net-BTP}
The Basic Transport Protocol (BTP) is a component of the ETSI specification.
It is a relatively simple protocol structure that varies depending on the scenario.
In the interactive scenario (BTP-A), the destination port and the source port (each field is 16 bits) are included.
In the non-interactive scenario (BTP-B), only the destination port and optional additional destination port information (each field is 16 bits) are provided.
BTP is specified in ETSI EN 302 636-5-1 \cite{EN302636-5-1} for \textit{Release 1} and in ETSI TS 103 836-5-1 \cite{TS103836-5-1} for \textit{Release 2}.
No sub-part 5-2 exists, and BTP remains the sole transport layer protocol.

The BTP port information is provided in ETSI TS 103 248 \cite{TS103248}.
As of 2024, 24 services are foreseen.
Two services (collective perception service and diagnosis, logging, and status service) are still under development.

The 'Amendments for LTE-V2X; Sub-part 2: Amendments to ETSI EN 302 636-5-1 Basic Transport Protocol (BTP)' \cite{TS102636-7-2} specifies the changes in regard to \cite{EN302636-5-1}.  
The changes are necessary because \cite{EN302636-5-1} assumed an interface-wise ITS-G5 with the required parameters, which are different for LTE-V2X.

\subsubsection[GN]{GeoNetworking (GN)}
\label{Chapter_ITSProtocols-TP-Net-GN}
The geographical networking, or short GeoNetworking (GN), is responsible for the distribution of messages over the different access layers, situated below the BTP.
To cope with this responsibility, GeoNetworking has media-dependent and media-independent specifications.
All of those specifications are included in the ETSI EN 302 636-X or ETSI TS 102 636-X standard families for \textit{Release 1}, as well as the TS 103 836-X family for \textit{Release 2}.
It should be noted that the TS designation in ETSI always begins with a 1, while the EN designation begins with a 3.
Some TS are upgraded to an EN to emphasize their role in the standardization process.

In \cite{EN302636-3} and \cite{TS103836-3}, the fundamental architecture of the GN network is delineated for \textit{Release 1} and \textit{Release 2}, respectively. 
The underlying requirements and scenarios are specified in \cite{EN302636-1} and \cite{EN302636-2}, respectively.
\cite{EN302636-3} delineates the architecture, components, and interfaces for GN.
The message formats and data structures, geographical addressing, forwarding mechanisms, and protocol operation are specified in Geographical addressing and forwarding for point-to-point and point-to-multipoint communications / Media-Independent Functionality (MIF) \cite{EN302636-4-1} for \textit{Release 1} and in \cite{TS103836-4-1} for \textit{Release 2}.
Despite the specification's media-independent designation, certain values are specific to CSMA/CA-based protocols like ITS-G5.
Accordingly, \cite{TS102636-7-1} stipulates an amendment with some interpretation for LTE-V2X.
The media-dependent specifications (Geographical addressing and forwarding for point-to-point and point-to-multipoint communications; Sub-part 2: Media-dependent functionalists (MDF)) exist as of 2024 for ITS-G5 \cite{TS102636-4-2} and LTE-V2X \cite{TS102636-4-3}.
Geographical addressing is based on the definition of different geographical areas, as specified in \cite{EN302931}.
These areas may be circular, rectangular, or multi-spared shapes.

The services transported inside of GN are identified by Application Object Identifiers (AIDs).
AIDs are used for worldwide service harmonization.
The current list is provided in \cite{TS102965}. 
The document is updated regularly to reflect new services.
The process of AID registration and handling is overseen by ISO as described in ISO 17419 \cite{EN17419}.
\cite{EN17419} in turn is an extension of ETSI TS 102 860 \cite{TS102860} from 2011.
Prior to 2016, ISO maintained a web version of the full list\footnote{The old list is still accessible via \url{https://standards.iso.org/iso/ts/17419/TS17419\%20Assigned\%20Numbers/TS17419_ITS-AID_AssignedNumbers.pdf}.}, but following this, no updates were published.

GeoNetworking is capable of transporting IPv6.
This is referred to as 'IPv6 Packets over GeoNetworking Protocols (GN6),' which is defined in \cite{EN302636-6-1} (\textit{Release 1}) and \cite{TS103836-6-1} (\textit{Release 2}).
Consequently, a sub-layer is introduced to map IPv6 functionalities onto GN.
This functionality is not used in deployment.

\subsubsection[WSMP]{WAVE Short Message Protocol (WSMP)}
\label{Chapter_ITSProtocols-TP-Net-WSMP}
The network and transport layer are combined in the networking services.
They are specified in the Wireless Access in Vehicular Environments (WAVE) specification, IEEE 1609-3 \cite{IEEE1609-3}, as part of the WAVE architecture \cite{IEEE1609-0}.
WAVE is designed to transport IPv6 and the Short Message Service.
In contrast, IP is only permitted for non-safety services on service channels.
The specifications delineate the data and management planes of the WAVE reference model, along with their service primitives.
The security plane is described in IEEE 1609-2 \cite{IEEE1609-2}.

WAVE defines two types of messages: WAVE Short Message (WSM) and WAVE Service Advertisement (WSA).
All information elements are defined in the common TLV (tag-length-value) format.
Services are identified by a Provider Service Identifier (PSID), which determines the service provided by a particular application.
Services are announced via WSA and consumed via WSM.
The PSID format (IEEE 1609-12 \cite{IEEE1609-12}) is related to ISO 17419 \cite{EN17419}.
A list of the currently registered services from SDOs, public organizations, and private companies is maintained by IEEE\footnote{PSID public list: \url{https://standards.ieee.org/products-programs/regauth/psid/public/}}.
In principle, all messages can be sent in a secured or unsecured manner, depending on the situation.
The security mechanisms are specified in IEEE 1609.2 \cite{IEEE1609-2}.

The WAVE Service Announcement (WSA) provides information about services provided by a particular station, including the PSID of the service and the channel on which the service is provided.
Additional information may include, for example, the IPv6 address, service ports, MAC addresses, and a Received Channel Power Indicator (RCPI) threshold, which must be met in order to consume a service.
Part of the WSA is the WAVE Routing Advertisement (WRA), which provides information about network connectivity.
This service allows stations to participate in a IPv6 network.
The message is derived from the router advertisement defined in RFC 4861 \cite{RFC4861}.

The WAVE Short Message Protocol (WSMP) is comprised of a network (WSMP-N) and a transport (WSMP-T) header.
Currently, WSMP-N specifies four message types: $\alpha$) \textit{Null-networking protocol} (standard), $\beta$) \textit{internal forwarding} within a station, $\gamma$) \textit{n-hop forwarding} broadcast forwarding over $n$ hops, and $\delta$) GeoNetworking as described in section \ref{Chapter_ITSProtocols-TP-Net-GN}.
Further extensions are possible.
The structure of WSMP-N is derived from the ISO CALM network protocol \cite{ISO29281-1}.
For the higher layer (WSMP-T), a Transport Protocol Identifier (TPID) is utlized.
The TPID currently specifies three cases (further extensions are possible): $\alpha$) PSID as the destination address without a source address, $\beta$) ITS port numbers \cite{ISO29281-1} as both the source and destination address, and $\gamma$) Local Port Protocol (LPP) (Japanese ARIB STD-T88, DSRC application sublayer).
The header may include an extension including the channel number, data rate, transmit power used, and channel load (all defined in \cite{IEEE802.11}).


\subsection{Direct communication}
\label{Chapter_ITSProtocols-TP-D}

\begin{figure*}[htbp]
    \centering
	\includegraphics[width=0.9\textwidth]{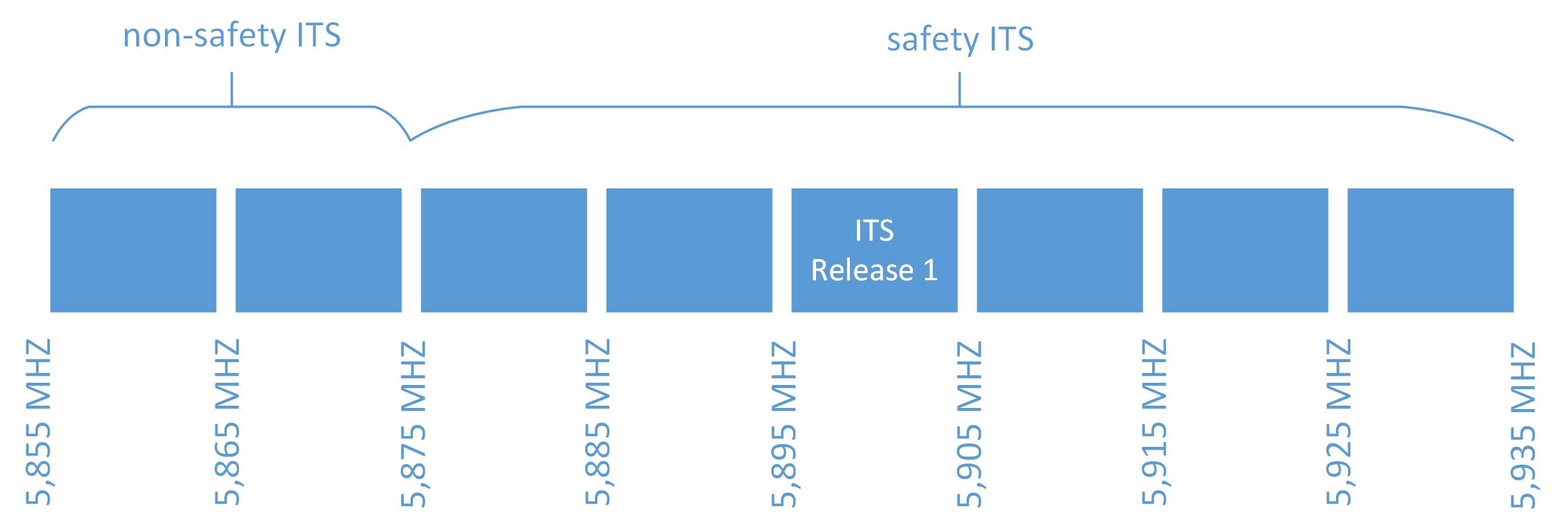}
    \caption{ITS spectrum}
    \label{fig_ITSSpectrum}
\end{figure*}

In the context of ITS, communication between systems can be either direct or infrastructure-based.
In Europe, direct communication is based on several specifications, which are outlined in ETSI TR 103 439 \cite{TR103439}.
This document also provides an overview of the specifications and their relations.
In principle, eight 10 MHz channels are available in the 5 GHz spectrum (see Figure \ref{fig_ITSSpectrum}).
Two of these channels are related to non-safety applications, while the reaming six are intended for safety-related applications.
In the upper two channels (between 5,915 MHz and 5,935 MHz), urban rail applications have priority.
The priority for the four channels between 5,875 MHz and 5,915 MHz is for road ITS applications.

The following list details some of the important regulations and requirements.
\begin{itemize}
	\item The \textit{harmonized channel specification} \cite{TS102724} provides channel usage guidelines regarding the use of ITS-G5 in the 5 GHz spectrum.
	\item ETSI ES 202 663 \cite{ES202663} established the European profile for layers one and two of ITS in the 5 GHz band. 
	\item The requirements for the equipment used in the 5,855 MHz to 5,925 MHz frequency band are specified in ETSI EN 302 571 \cite{EN302571}.
	\item It is imperative that ITS equipment in the 5 GHz band not impede the functionality of existing applications. In order to facilitate tolling, protection zones \cite{TS102792} were specified that can be transmitted via direct communication (see section \ref{Chapter_ITSProtocols-TP-ITSApp-IMZM}) or, at least in the case of tolling, can be pre-installed in a station (including regular updates).
	\item In order to accommodate the priority in the urban rail spectrum part, measures must be taken. An example would be the CSMA/CA of IEEE 802.11.
\end{itemize}

In Europe, spectrum regulations impose several requirements for usage.
Two of these are technology neutrality and backward compatibility.
Technology neutrality signifies that, from a regulatory standpoint, no technology should be hindered or favored. 
It does not imply that one technology cannot be selected for a particular spectrum.
For instance, the spectrum for cellular technology is clearly defined.
This is also true for ITS applications, given that the spectrum is only granted for ITS applications, but no specific access technology.
Backward compatibility implies that systems and services already in the market should be able to serve their intended purpose.
New systems should be able to interact with those legacy systems.
However, the legacy system should also not hinder the development of new and enhanced technologies.
According to the European Commission, nearly one million vehicles using the ITS-G5 and several thousand infrastructure systems are deployed \cite{C2CCCForum2022}\footnote{Additional information: \url{https://www.car-2-car.org/fileadmin/documents/General_Documents/C2C-CC_Statement_-_ITS_Directive.pdf}}.

The ITS band can be utilized with a variety of technologies.
These technologies were developed by different groups with the same objective in mind.
Currently, there is a focus on ensuring the interoperability of these technologies (e.g., \cite{TR103576-2}).
In Europe, the ITS-G5 \cite{EN302663} and LTE-V2X \cite{EN303613} specifications are in place, with further advancements anticipated.
Long-range technologies such as cellular are not permitted on that frequency.
They are discussed in section \ref{Chapter_ITSProtocols-TP-I}.

It should be noted that ITS-G5 and LTE-V2X are interoperable at the lower layers.
This means that either technology can receive or send the data of the other.
Both access technologies are currently in the process of development.
The evolution of ITS-G5/IEEE802.11p will be IEEE802.11bd.
The goal of this development was achieve interoperability and backward compatibility with the former version.
The evolution of LTE-V2X will be 5G-V2X.
Currently, both are not compatible with each other on the physical layer.
In ETSI TR 103 576-2 \cite{TR103576-2}, some considerations and techniques are presented which describe how different access technologies could be interoperable.
The Car 2 Car Communication consortium published two white papers (\cite{C2CCCwpCoexsistance2021}, \cite{C2CCCwpCoexsistance2022}) on the subject of coexistence and mitigation between ITS-G5 and LTE-V2X.

\subsubsection[DSRC]{Dedicated short-range communication (DSRC)}
\label{Chapter_ITSProtocols-TP-D-DSRC}
Dedicated Short-Range Communication (DSRC) encompasses a range of technologies, including dedicated radio-based tolling and Vehicular Ad-Hoc Networks (VANETs).
In this paper, the latter will be the focus of our attention.

The ITS-G5 \cite{EN302663} is the ETSI specification for the physical and data-link layer communication technology based on WLAN.
In 2010, IEEE specified the IEEE 802.11p (\cite{IEEE802.11p}) standard.
It is an extension of the IEEE 802.11 specifications and is now part of the current IEEE 802.11-2020 (\cite{IEEE802.11}) specification from 2021.
It describes the settings for direct communication (e.g., outside the context of a basic service set, no channel scanning, channel coding, etc.). 
The specification is based on the IEEE 802.11a specification from 1999.
To meet the demands of modern applications and regulations (e.g., spectral efficiency), the IEEE 802.11bd \cite{IEEE802.11bd} specification, also known as "next-generation V2X" (NGV) was ratified in 2022. 
This version will be included in the next release of the 802.11 specifications.
For example, an overview of new features is provided in \cite{V2XnextGen2019}.

It is worth noting that the communication range is, based on the results of the simTD\footnote{More information about the project: \url{https://www.simtd.de}} field trial,  between 250 m (in urban areas) and 450 m (on motorway) under normal conditions.
However, under ideal conditions, the communication range can be up to 6,000 meters \cite{PetryMsc2013}.

To coordinate the different communication participants, a feature called \textit{Decentralized Congestion Control} (DCC) is implemented \cite{TS103175}, \cite{TS102687}.
DCC is necessary to address the dynamic nature of the communication environment, including physical condition and the number of participants, as well as the specific needs of the applications.
To this end, DCC is active at the access and data link layers and considers the applications in operation.
With this data, it determines the optimal time and type of message to be sent.

\subsubsection[C-V2X]{Cellular Vehicle-to-everything (C-V2X)}
\label{Chapter_ITSProtocols-TP-D-CV2X}
The second solution is based on cellular technology: Cellular Vehicle-to-Everything (C-V2X) communication.
Initially, this was specified for the fourth generation of cellular technology, designated as LTE-V2X in ETSI EN 303 613 (\cite{EN303613} based on \cite{TS103613}).
The next generation, based on 5G new radio (NR-V2X), has already been standardized \cite{TS124587}.
Extensive tutorials about the LTE-V2X and NR-V2X can be found in \cite{Garcia2021} and \cite{Gonzalez2022}.

Both can transport data via the two interfaces PC5 and Uu \cite{TS124587}.
PC5 is the interface for direct communication between systems.
This can be achieved either with infrastructure coordination (mode 3) or without infrastructure coordination (mode 4).
Mode 4 is the most comparable to ITS-G5.
An introduction to mode 4 can be found in \cite{Molina2015}.
The PC5 interface can transport V2X messages directly, IP and non-IP based V2X messages, and IPv6 (no IPv4) within V2X messages.
The Uu interface is employed for communication between stations and a V2X application server situated within the cellular core network.
The interface is capable of transporting plain V2X messages, IP and non-IP based V2X messages, V2X messages as payload in UDP/IP or TCP/IP, and V2X messages as part of the user data.
All messages are transmitted as unicast messages.
No broadcast (e.g., eMBMS) transmission is supported.

Currently, congestion control is standardized for LTE-V2X only \cite{TS103574}.
It only applies to the PC5 interface in mode 4.
Its focus is on the ITS access layer (access and data link layer) with input from the ITS networking and transport and the ITS facilities layers.

\subsubsection{Other protocols}
\label{Chapter_ITSProtocols-TP-D-Oth}
The two technologies mentioned above are the dominant ones.
However, other technologies can be used in different scenarios or with varying application requirements.
A selection of these technologies is described below.

\paragraph{Bluetooth}
\label{Chapter_ITSProtocols-TP-D-Oth-Bt}
Bluetooth is a short-range communication technology for integrated devices with low power requirements.
The specification was maintained by IEEE until version 2.1 in IEEE 802.15.1. 
The most recent version of this specification is from 2005 and was deactivated in 2018.
New versions are developed by the Bluetooth Special Interest Group\footnote{More information about Bluetooth: \url{https://www.bluetooth.com/}}
The current version is 5.3 \cite{BluetoothCoreSpec}.

In the field of ITS, Bluetooth is employed, for instance, for vehicle counting.
Active Bluetooth devices transmit on-demand information about themselves.
This can be used to count vehicles at intersections or other traffic infrastructure systems (realized as so-called \textit{Bluetooth Low Enebeacons}).
Bluetooth systems typically send the same information, which is used to track the travel time of vehicles through a specific street segment.

Bluetooth operates in the 2.4 GHz \textit{Industrial, Scientific, and Medical} (ISM) band, alongside other applications in this band.
In principle, Bluetooth operates in a connection-oriented manner, whereby Bluetooth low energy supports connection-less communication.
Every device can act as both a service provider and a service consumer.
Bluetooth employs profiles, which represent protocol specifications for various applications, including audio, video, telephone, TCP/IP, and Wireless Application Protocol (WAP).
The communication range is typically limited to 100 meters.
Bluetooth is capable of utilizing a variety of physical and data link (only media access control part) layer technologies, including IEEE 802.11.

\paragraph{RFID}
\label{Chapter_ITSProtocols-TP-Oth0-D-RFID}
Radio-frequency identification (RFID) is a technology that operates within an electromagnetic field.
Its primary function is to track and identify any object or human being. 
There is no internationally harmonized specification for RFID.
Instead, several specifications exist from organizations such as ISO or IEC.
Furthermore, the frequency band is not harmonized globally. 
Only the low- and high-frequency bands (25–134.2 kHz, 140–148.5 kHz, and 13.56 MHz) are included in the ISM band and, therefore, usable worldwide.
The other band in ultra-high frequency, microwave, and mm-wave are not harmonized.

RFID operates with two primary system components: the tag and the reader.
The tag is a small, often flexible device that holds the information.
The reader reads the information from the tag.
Tags and readers can be active or passive, with or without a power supply.
The most common scenario is that the tag is passive, and the reader is active.
This means that the reader induces power into the tag, which enables the tag to send its information to the reader.
The range of an RFID transmission depends on the frequency and the active or passive status.
In the typical case described above, a few centimeters up to a few meters (theoretically up to 100 m) are possible.

Currently, RFID is used in different fields of ITS.
In logistics, it is used to track items around the global supply chain.
It is also used to authenticate vehicles to access a specific area or for tolling.
Another angle would be the identification of a vehicle system also for payment at, e.g., a charging station.

\paragraph{Millimeter-Wave}
\label{Chapter_ITSProtocols-TP-Oth-D-MmWave}
The communication of massive data (e.g., LIDAR sensor data sharing) with a short delay may be required for safe automated driving.
Since 2008, numerous researchers have conducted simulations and experiments to use millimeter-wave communication in different parts of the 28 GHz to 60 GHz spectrum.
The 28 GHz to 60 GHz spectrum has the advantage of broad channels and can support a bandwidth of over 1 gigabit per second with a delay of less than 10 ms (\cite{V2XnextGen2019}).
However, the high spectrum leads to a limited communication range of typically less than 100 meters in static and even less in dynamic scenarios.
Additionally, line-of-side communication is required, and nearly any obstacle can reduce the communication range and increase the packet error rate.
For instance, \cite{MmWaveV2X} employs a communication range of 5 meters for practical assessment, yet assert a static use case might permit a range of up to 200 meters. 
In this instance, consumer electronics were utilized, and no specialized V2X technology.
The technology is currently undergoing standardization and will be available for specific use cases.
It is not intended to replace dedicated safety communication technologies, but rather to complement them (see chapters \ref{Chapter_ITSProtocols-TP-D-DSRC}, \ref{Chapter_ITSProtocols-TP-D-CV2X}) with sensor sharing capacity.

\paragraph{Vehicular Visible Light Communications}
\label{Chapter_ITSProtocols-TP-Oth-D-VVLC}
In addition to communication based on radio frequency technologies, other technologies are currently under discussion.
One example is vehicular visible light communication (V-VLC), which is explained, for example, in \cite{VVLC2019} and \cite{VVLC2021}.
The current development of V-VLC is standardized by the IEEE P802.11 Light Communication (LC) task group as IEEE 802.11bb \cite{IEEE802.11bb}. 
In brief, the specification is an extension to Wireless LAN that modifies the physical and data-link layer for the usage of visible light.

It is worth noting that the IEEE 802.11bb standard bears a resemblance to the IEEE 802.11bd standard, which is discussed in chapter \ref{Chapter_ITSProtocols-TP-D-DSRC}.
Both standards are distinct and should not be confused.

The objective of V-VLC is to utilize visible light based on light-emitting diodes (LEDs) to facilitate vehicle communication. 
The sender could employ, for instance, the existing head and tail lights for this type of communication.
The receiver could be an existing camera (e.g., currently used for traffic monitoring (ADAS) or sign detection) or photodiode. 
The reuse of existing or commercial off-the-shelf (COST) products would allow for a very low-cost solution.
However, communication is still in development and faces some challenges.
Currently, line-of-sight communication is possible and foreseen.
V-VLC has a high directionality, resulting in a minor collision domain. 
This allows for high spatial reuse of the modulation bandwidth for devices in close proximity (\cite{VVLC2021}).
Potential security breaches can be easily identified because known attacks are based on obstruction.
However, it is possible that future attacks may emerge.

A variety of modulation and coding schemata were subjected to rigorous evaluation in real-world experiments \cite{VVLC2021}.
The tests were conducted under both daylight and nighttime conditions.
In the sunlight, a stable data rate of 1.35 Mbit/s could be achieved over a distance of up to 40 meters in a steady scenario with a 64-QAM modulation and a coding schema of 3/4.
A maximum data rate of 8.8 Mbit/s was measured between two vehicles positioned 5 meters apart.
In realistic traffic scenarios and distances between vehicles on urban roads, a data rate of 0.15 Mbit/s to 0.9 Mbit/s could be achieved.
The maximum distance currently possible is 70 meters.
However, this distance could be extended with more sophisticated sender and receiver equipment.

One challenge in visual light communication is the potential for interference from other light sources, such as the sun or halogen lights.
This is particularly problematic when the unwanted light directly shines inside the receiving sensor, such as a camera or photodiode.


\subsection{Infrastructure communication}
\label{Chapter_ITSProtocols-TP-I}

Communication in ITS depends on the information available via infrastructure communication.
Examples of such information include traffic information, weather information, logistic information, map updates, and do forth.
There are different kinds of infrastructure systems.
One example is cellular (see section \ref{Chapter_ITSProtocols-TP-I-Cel}) and the other is Digital Audio Broadcast (DAB) (see section \ref{Chapter_ITSProtocols-TP-I-DAB}).

\subsubsection{Cellular protocols}
\label{Chapter_ITSProtocols-TP-I-Cel}
Cellular networks are commonly used in the lives of many people.
With different generations currently on the market, cellular technology can provide service for different scenarios.
3G (\cite{TS125300}), 4G/LTE (\cite{TS123401}, \cite{TS136300}, and 5G \cite{TS138300} can provide data transfer of any kind.
Techniques such as mobile edge computing and network slicing may facilitate an agile system design and allow for time-critical computation and information distribution at the network's edge.
However, techniques such as fog computing (see \cite{Bonomi2012}) are not deployable without mobile edge clouds.
The concept of the 'tactile internet' \cite{Fettweis2014} posits that cellular networks can provide a delay as low as 1 ms from one terminal to another via the core networks. 
Even 5G does not achieve this level of performance \cite{Rischke2021}, with the next generation 6G expected to further reduce the time needed.

The use of cellular services for ITS is extensively described in \cite{TR102962}.
One aspect is the so-called \textit{GeoServer} for the geographical distribution of ITS-related messages.
This concept was developed interdependently by Ericsson (variable-sized rectangular-grid based) and the author organization (variable geographical area based) around 2010.
With these solutions, a service provider can use the infrastructure to distribute information in a specific geographical area.

Cellular networks can distribute information in a manner that differs from a traditional one-to-one relationship.
Instead, they employ broadcast and multicast techniques.
This approach is known as the Multimedia Broadcast Multicast Service (MBMS), encompasses the Stage 1 \cite{TS122146}, Architecture \cite{TS123246}, and protocols \cite{TS126346} specifications.
Its subsequent evolution, the Evolved Multimedia Broadcast Multicast Service (eMBMS), is outlined in eMBMS \cite[Chapter 15]{TS136300} document.
The initial design of the system was for the distribution of multimedia content, such as video streaming.
However, it is now possible to distribute ITS-related information in some geographical regions \cite{TR102962}.

\subsubsection[DAB]{Digital Audio Broadcasting (DAB)}
\label{Chapter_ITSProtocols-TP-I-DAB}
Digital Audio Broadcasting (DAB) \cite{EN300401} is a communication technology designed to provide audio and data services to users.
It was developed by the EU-funded Eureka Project 147 and was first published in 2000.
The underlying concept was the digitization of the terrestrial radio audio service.
The data is provided and broadcast in a unidirectional manner from the sender to the receiver, without any back channel.
The majority of large-scale transmission towers are utilized to disseminate the information over a vast geographical area.
DAB is capable of theoretically covering a radius of up to 96 km with a net data rate of 2,304 kbit/s in the very high frequency (VHF) bands I, II, and III (between 47 MHz and 230 MHz). 
DAB has the advantage that it can be deployed in a so-called \textit{single frequency network} (SFN) in which all transmitters use the same frequency to cover a large area.
In addition to DAB, there is DAB+ \cite{TS102563}.
The data coding mechanisms, energy consumption, and spectral efficiency differ between the two.

Service in DAB can be transmitted encrypted and unencrypted, which allows the provision of private or paid services.
There are a few data services specified for DAB.
\begin{itemize}
	\item The Multimedia Object Transfer Protocol (MOT) \cite{EN301234} and Dynamic Label Segment(+) (DLS) \cite{EN300401}, \cite{TS102980} are services for the transmission of supplementary audio information, such as song names and interpreters, short news items, and small images or logos.
	\item Journaline \cite{TS102979} is a hierarchical text message service that is employed, for instance, in the Emergency Warning Functionality (EWF) as part of the German modular warning system (Modulares Warnsystem, MoWaS).
	\item The Traffic Message Channel (TMC) and its successor TPEG (see section \ref{Chapter_ITSProtocols-TP-PS-TPEG2}) are utilized for the distribution of traffic-related information (e.g., routing information, traffic jams, accidents, etc.).
	\item It is possible to transmit IP packets via DAB using IP over DAB \cite{TS101735}. However, it is important to note that the absence of a back channel must be considered, and only connectionless and unacknowledged services may be used.
\end{itemize}

\section{ITS communication architectures and data exchange platforms}
\label{Chapter_ITSProtocols-ConnArchDataExchange}

From the perspective of communication, ITS may be regarded as a data distribution system.
Such a system necessitates an architecture that enables the retrieval of any information required to enhance traffic safety and efficiency.
Various approaches from research and public authorities describe the ITS architecture.
This section presents three of these approaches: the German research project Converge, the two European projects NordicWay and InterCor, and the C-Roads architecture, which exemplifies the public authorities' perspective.

Subsequently, examples of data exchange platforms are presented.

\subsection{CONVERGE} 
\label{Chapter_ITSProtocols-ConnArchDataExchange-CONVERGE}
In the German research project CONVERGE\footnote{CONVERGE project \url{https://fgvt.htwsaar.de/site/en/converge/}} (2012-2015), an open architecture was developed \cite{Vogt2013}, \cite{Wieker2014}.
This architecture is based on the so-called system's network.
The network is comprised of three layers: $\alpha$) the access layer, which includes people, vehicles, and traffic infrastructure systems; $\beta$) the network layer, which encompasses various communication technologies; and $\gamma$) the backend layer, which houses the service provider, TMCs, and support services (e.g., security providers, service directory, etc.).
On top of the technical systems, a management layer is introduced.
This layer represents the governance structure necessary to introduce rules and regulations, rules enforcement, and top-level security functions.
A role-based model accompanies the architecture, including economical, technical, and institutional roles.
These roles serve as the basic principle for a robust and economically feasible operation.

\subsection{NordicWay} 
\label{Chapter_ITSProtocols-ConnArchDataExchange-NordicWay}

The NordicWay EU-funded projects 1, 2, and 3\footnote{NordicWay \url{https://www.nordicway.net/}} developed an architecture for cross-border and cross-domain data transfer \cite{nordicwayD21.1} \cite{nordicwayArch}.
This architecture is focused on the realization of use cases and is based on a service provider concept.
The service providers are the backend systems (road operators, content providers, application providers, and the interchange system), the road user systems (vehicles), and the roadside systems (traffic signal and detector systems).
The establishment of 'interchange systems' serves to reduce the overall traffic and the service workload.
They act as a proxy between the two regions and provide control information about the supported technical capabilities (e.g., protocol format), the country, and the supplied or wanted information.
The same interface is used for the actual data transfer.

\subsection{InterCor} 
\label{Chapter_ITSProtocols-ConnArchDataExchange-InterCor}
The objective of the Interoperable Corridors Deploying Cooperative Intelligent Transport Systems project\footnote{InterCor \url{https://intercor-project.eu/}} (InterCor) was to develop an architectural framework for connecting disparate corridors across Europe.
The core component of this fundamental architecture is the hybrid communication approach (\cite{intercorM4v2.1}, \cite{intercorM4Hy}, \cite{Crockford2018}), which specifies three interfaces for interoperability: IF 1 (ITS-G5 air interface), IF 2 (back-office interface), and IF 3 (cellular interface).
The IF 2 provides connectivity between different countries.
This is necessary due to the fact that countries have different regulations and legal frameworks.
Via this interface, a vehicle $V$ from country $A$ can access a service $N$ in country $B$ in a manner similar to accessing a service in its home country.
The information is transmitted via IF 1 or over the backend to the service provider in country $A$, and then via cellular (IF 3) to vehicle $V$.

\subsection{C-Roads} 
\label{Chapter_ITSProtocols-ConnArchDataExchange-CRoads}
The C-Roads Platform is a collaborative organization of road operators and public authorities.
Its objective is the deployment of C-ITS in Europe.
Harmonizing the deployment of C-ITS was deemed necessary to reach this goal.
Therefore, a set of profile specifications \cite{CROADSHCP} was released and is continuously updated.
The profile documents contain information about infrastructure-related services (e.g., traffic light-related and warning messages), RSU behavior, system management, security, and (cross) border testing.
The C-Roads architecture comprises four connections: $i$) backend to backend, $ii$) backend to ITS-G5 roadside infrastructure, $iii$) ITS-G5 roadside infrastructure to end user, and $iv$) end user to backend.
The connections $ii$ and $iv$ are not specified and may or may not conform to the \textit{EU C-ITS Trust Domain}.
The trust domain is an organizational domain where all stakeholders and participants follow the same set of security rules and provide services with the expected level of reliability.
Interfaces $i$ and $iii$ are located within that trust domain.
The interface $iii$ is based on the ITS-G5 mechanisms described in chapter \ref{Chapter_ITSProtocols-TP-D-CV2X}.
The interface $i$ can be either basic or improved.
The basic interface employs C-ITS messages (see chapters \ref{Chapter_ITSProtocols-TP-ITSApp}, \ref{Chapter_ITSProtocols-TP-ITSApp-secp}, \ref{Chapter_ITSProtocols-TP-Net-BTP}, and \ref{Chapter_ITSProtocols-TP-Net-GN}) based on an AMQP network (see chapter \ref{Chapter_ITSProtocols-CP-IP-AMQP}).
The improved interface utilizes the basic interface messages and architecture, but it employs the interchange systems from NordicWay (see chapter \ref{Chapter_ITSProtocols-ConnArchDataExchange-NordicWay}) between regions or countries.

\subsection{Data exchange platforms} 
\label{Chapter_ITSProtocols-ConnArchDataExchange-DataExchangePlatforms}

In the European Union, the necessity for mobility and transport data exchange is regarded as a pivotal element in the advancement of joint European development.
In accordance with the delegated acts of the ITS-Directive (2010/40/EU)\footnote{Directive 2010/40/EU of the European Parliament and of the Council of 7 July 2010 on the framework for the deployment of Intelligent Transport Systems in the field of road transport and for interfaces with other modes of transport Text with EEA relevance, \url{https://eur-lex.europa.eu/legal-content/EN/TXT/?uri=CELEX\%3A32010L0040}}, each member state is required to establish a National Access Point (NAP)\footnote{More information about the national access points and the delegate acts are available online \url{https://transport.ec.europa.eu/transport-themes/intelligent-transport-systems/road/action-plan-and-directive/national-access-points_en}} to serve as a singular entity for traffic and transport-related data. 
Currently, only a portion of the already digitized data is available through the portal.
The objective is to include all digitally available data and digitize all available traffic data in this context.
The information exchange should be based on DATEX2.
The EU established the National Access Point Coordination Organisation for Europe (NAPCORE)\footnote{NAPCORE: \url{https://napcore.eu/}} to coordinate the mobility platforms.
Public data should be available freely, but regulations apply for private data included in the national access points.
It is possible to restrict the usage of the data to specific purposes.
The current list of national access points can be found online\footnote{European Commission, DG Transport: \url{https://transport.ec.europa.eu/document/download/963c997d-efd9-40ae-a38b-5d4b935bdfcf_en?filename=its-national-access-points-2023-09-19.pdf}}.
Examples of NAPs include the mobilithek\footnote{Mobilithek: \url{https://mobilithek.info/}} and the Slovakia Odoprave \footnote{Odoprave: \url{https://odoprave.info/}}.
Private exchange systems like the TLEX platform \footnote{TLEX I2V PLATTFORM: \url{https://www.monotch.com/tlex-i2v/}} can provide any data related to transport on their own terms.

\section{Detailed Traffic-Related Specifications} 
\label{Appendix_PE}

This section offers further insights into a selection of traffic and communication-related ITS protocols.
The four traffic-related specifications are RiLSA, TS 19091, TPEG2, and Vehicle-2-Grid.
A communication-related protocol is the service announcement.


\subsection{RiLSA}
\label{Appendix_PE-TP-RiLSA}

The 2015 edition is comprised of nine sections.
\begin{enumerate}
	\item \textbf{Basics} The initial chapter is devoted to the regulation and guidelines for the installation and operation of traffic lights, including an examination of various types of traffic lights for both vehicles and trains.
	\item \textbf{Design of the signal program} A signal program determines how the traffic is routed through an intersection based on various factors, including the view of all streets and lanes, including vehicles, buses, bicycles, crossings, parking spots, channeling islands, and traffic lights. Additionally, it considers statistics on vehicle density based on HBS and traffic accidents. Several safety concerns regarding turning, channeling islands, green filter arrows, streetcars, buses, pedestrians, and cyclists are described. The specification includes examples of phase and time-distance diagrams to illustrate the rules. It is important to consider the number of phases, their sequence, the involved signal group, and phase transitions. To ensure safety, various timers must be calculated, including the duration of the yellow light, the delay between a red light and green light for crossing traffic, the duration of the green light, and the time it takes for the traffic light to return to the original state. These timers are dependent on the type of vehicles involved, including bicycles. The aforementioned information is summarized in a signal time plan.
	\item \textbf{Interaction between signal light control and the design of the road traffic systems} In addition to technological aspects and program logic, it is important to consider the physical road. This includes the direction, number, type, and size of lanes (including left or right turns and straight through), bicycle traffic, median strips, traffic islands, roadbed crossings, and pedestrian and cyclist crossings, as well as bus stops.
	\item \textbf{Control procedures} The control procedures can be divided into two parts:  macroscopic and microscopic. The macroscopic procedures involve long-term or traffic-dependent signal programs and suitable framework programs, which provide the framework for the microscopic programs. The latter include short-term changes in traffic flow and state, typically lasting a few seconds or the signal period. The system takes into account the number of vehicles waiting at a traffic light (detected, e.g., through cameras or detection loops), requests from public transport vehicles, and pedestrians. The coordination of traffic lights can be done locally and globally. In the global case, more than one traffic light can be coordinated. Public transport vehicles are also considered when coordinating multiple traffic lights for a green wave.
	\item \textbf{Special types of signaling} In certain situations, a regular traffic light crossing may not be provided, and instead, special road conditions must be taken into account. These are situations where traffic lights are not fully installed, such as to improve safety or reduce wait time. Other examples include signaling to indicate road closures due to construction, lane direction changes for events such as fairs or concerts, and regulations for entering highways.
	\item \textbf{Technical execution} The traffic light system is composed of various components, including a control system for switching logic, signal heads for vehicles, pedestrians, public transport, and cyclists, systems for people with disabilities, signs, detectors, and structural conditions. 
	\item \textbf{Technical approval and operation} Safety is of utmost importance in traffic. Therefore, strict safety tests must be conducted and safety rules must be followed before a traffic light is allowed to operate. The circumstances in which safety measures are required are clearly defined, and they may be either remissible, conditionally remissible, or obligatory. Additionally, guidelines for the use of signs in the event of a power failure or other disturbances, as well as instructions for police intervention in the current traffic light program, must be established.
	\item \textbf{Quality management}  Quality management involves design and implementation of necessary documentation in all phases, including planning and operations management (OM). It also includes scheduling reevaluation and testing procedures.
	\item \textbf{Related rules and standards} \cite{Ril15} mentions 23 standards, 25 recommendations, and eight laws or bylaws.
\end{enumerate}

The RiLSA \cite{Ril15} is not primarily a communication-related guideline. 
Nevertheless, it does describe communication requirements and several communication relations without delving into detail or linking to a suitable standard. 
The following paragraphs list the communication-related aspects.

\textbf{Section 4.2.2.2} (Request for a green period) and \textbf{section 6.2.10} (Signal heads for public transport vehicles) require the detection of public transport or emergency vehicles through announcements and a notice of departure. 
The message should include the type of vehicle, line number, and course, depending on the vehicle type. 
No protocols, messages, or data formats are specified for this purpose, nor are any timing or other constraints mentioned. 
Section 5.1.3.2 (Public transport acceleration and prioritization) outlines the communication necessities for a variety of physical junction and roundabout scenarios.

\textbf{Sections 4.3.1.1} (Framework conditions for signal programs), \textbf{4.4} (Coordination for signal control procedures), and \textbf{6.1} (Traffic light controller) outline the prerequisites for a traffic light controller that can coordinate multiple traffic lights through network controls. 
The objective of this coordination is to create a green wave for all traffic participants.

The opening of a lane (as described in \textbf{section 5.3.4}, Change of operating state) is only permitted if the lanes can be controlled and are free of standing vehicles (in the case of a hard shoulder) or opposing traffic (in the case of dynamic lane allocation with opposing traffic). 
If such a problem is detected, the lane must be closed. 
The text only defined the transport of information without providing exact measurements or constraints.

\textbf{Section 6.3} (Detection systems) specifies several measurable data elements and the corresponding derived values. 
For further information, please refer to \cite{FGSV312}. 

For temporal traffic lights at roadworks (\textbf{section 7.4.3}, Alternative signal control), communication can be achieved through cable or wireless means. 
The specification does not provide any additional information on the available technologies or protocols. 
It is necessary to comply with \cite{EN50556}, which covers electronics and other hardware safety aspects for traffic lights.

Quality management (\textbf{section 8.3}, \textit{Preconditions}) requires precise documentation of  data flow, interfaces, and data consistency. 
However, it does not specify the methods or requirements for fulfilling this. 
Additionally, \textbf{section 8.6.2} (\textit{Quality analysis at junctions}) states that all traffic and process-related data is stored and can be used for analysis in a traffic-dependent system configuration.

In conclusion, RiLSA \cite{Ril15} is not a communication-related guideline. 
THe standard describes high-level communication requirements and relations without providing detailed information. 
Although the standard is not intended for a communication system, it addresses many relations, and therefore, more information should have been provided. 
It is not possible to extract functional requirements for a communication system.

\subsection{TS 19091}
\label{Appendix_PE-TP-TS19091}

The 2019 version of the standard comprises eight parts and seven appendices. 
\begin{enumerate}
	\item[1-4] The initial four chapters cover the \textbf{Scope}, \textbf{Normative references}, \textbf{Terms and definitions}, and \textbf{Abbreviated terms}. 
	\item[5] \textbf{General description} The communication systems and data elements are based on three categories of use cases:
		\begin{itemize}
			\item \textbf{Safety use cases} The primary objective of safety use cases is to provide approaching vehicles with information regarding the intersection, whether signalized or non-signalized, with MAP and SPaT messages. This information includes permitted maneuvers and the time when they will change. If it is detected that a vehicle will not be able to stop in time, the data from the vehicle could be used to alter the traffic light's behavior. The the event that no change is possible, a warning message will be disseminated to all approaching vehicles. 
			\item \textbf{Mobility/Sustainability use cases} These use cases aim to improve traffic efficiency by reducing environmental impact and improving traffic flow. Vehicles are provided with information about signal phases and upcoming changes so that they can adjust their speed accordingly. The platooning case is mentioned but will be considered in a future version of the standard. Payment options for these services are noted as out of scope for the standard. 
			\item \textbf{Priority/pre-emption use cases} Cities frequently aim to provide efficient and appealing public transportation options for their citizens. As a result, public transport vehicles may request priority service at intersections. Additionally, the ability for heavy-duty trucks or a platoon of trucks to navigate through a city is also a relevant use case. However, the most critical consideration is ensuring clear and safe routes for emergency and public safety vehicles, such as fire brigades, ambulances, and police. It is possible for vehicles to request a green light at an intersection via an SRM. This is done by sending a message to the traffic light controller, which can include the vehicle's speed and heading. Based on the MAP message, a path through the intersection is requested. The traffic light controller responds with an SSM message indicating the type of service provided. The changes in the traffic light are visible through the SPaT message. 
		\end{itemize}
		The functional model is based on three components: vehicle, roadside infrastructure, and traffic/fleet management centers. The vehicle should be able to handle the following messages: CAM (send direction, see \ref{Chapter_ITSProtocols-TP-ITSApp-CAM}), MAP (receive direction, see \ref{Chapter_ITSProtocols-TP-ITSApp-MAP}), SRM (send direction, see \ref{Chapter_ITSProtocols-TP-ITSApp-SRM}), SPaT (receive direction, see \ref{Chapter_ITSProtocols-TP-ITSApp-SPaT}), and SSM (receive direction, see \ref{Chapter_ITSProtocols-TP-ITSApp-SSM}). The roadside equipment (RSU) should be capable of bidirectional message handling. It should also be able to communicate with both the TLC and the TMC. The TMC manages traffic-related information, other backend services, and connections to other ITS services. Additionally, some basic assumptions regarding the use cases are defined.
	\item[6] \textbf{Functional description} This section outlines the prerequisites for implementing the system and the use cases. It encompasses communication solutions and bandwidth, as well as media constraints. It is assumed that BSM, CAM, or DENM will be distributed for every connected vehicle. The chapter is divided into the following sections: 'public safety vehicles', 'signal pre-emption', 'public transport and commercial vehicles', 'signal priority requirements', 'broadcast area's geometry', 'broadcast GNSS augmentation details', 'signalized intersection requirements', 'broadcast cross-traffic information', 'broadcast vulnerable road user sensor information', 'broadcast dilemma zone violation warning', 'broadcast signal preferential treatment status', 'message identifier', 'system performance requirements', 'transition rates - signal preferential treatment', 'transmission rate requirements - broadcast roadway geometrics information', 'transmission rate requirements - GNSS augmentations detail broadcast', 'transmission rate requirements - broadcast signal phase and timing information', 'transmission rate requirements - broadcast cross-traffic sensor information', and 'transmission rate requirements - broadcast vulnerable road user sensor information'.
	\item[7] \textbf{Messages} This chapter presents a list of four messages for which data dictionaries have been defined in this standard for international deployment. These include the messages presented in chapters \ref{Chapter_ITSProtocols-TP-ITSApp-MAP}, \ref{Chapter_ITSProtocols-TP-ITSApp-SPaT}, \ref{Chapter_ITSProtocols-TP-ITSApp-SRM}, and \ref{Chapter_ITSProtocols-TP-ITSApp-SSM}. Annexes E, F, and G define the profiles A, B, and C for \cite{SAEJ2735}.
	\item[8] \textbf{Conformance} The conformance chapter identifies three conditions that must be met for an implementation to be considered in accordance with the standard.
		\begin{itemize}
			\item The data content must fulfill the mandatory and \textit{selected} optional requirements for the requirements traceability matrix and the relevant annexes.
			\item The message structure must adhere to the selected annex.
			\item The system interfaces must implement all data elements from the requirements but only use those messages and data elements that a system must support.
		\end{itemize}
	\item[A] \textbf{Annex A} serves as a baseline for the standard, delineating the use cases in question. These include nine priority use cases, seven safety use cases, and nine mobility use cases. Every use case description is composed of the following elements: name, category (mobility, safety, sustainability), infrastructure role (involved systems, communication directions), brief description, objective, constraints (communication technologies that can be used), geographic scope, actors, illustration (use case and parts of the communication), preconditions, main flow, alternate flow(s), post-conditions, information requirements (the messages and data fields of the messages that are used in the use case), issues (assumptions), and source documents/references. 
	\item[B] \textbf{Annex B} presents a matrix mapping of all use cases with all requirements. It specifies whether the requirements are mandatory, optional, conditional, excluded, or otherwise linked to the use cases.
	\item[C] \textbf{Annex C} is a matrix that maps requirements to the messages SPaT, MAP, SSM, SRM, and RTCM, including the fields in the message, if applicable. 
	\item[D] \textbf{Annex D} delineates the procedure for extending the SAE J2735 \cite{SAEJ2735} regarding regional extensions for SPaT, MAP, SSM, and SRM.
	\item[E] \textbf{Annex E} presents Profile A of \cite{SAEJ2735} for North America.
	\item[F] \textbf{Annex F} presents Profile B of \cite{SAEJ2735} for Japan.
	\item[G] \textbf{Annex G} presents Profile C of \cite{SAEJ2735} for Europe. including data elements specified in \cite{TS102894-2}.
\end{enumerate}

The initial four sections lack pertinent data for the present analysis. 

\textbf{Section 5.2.1} (\textit{Description} in section Functional model) states that the focus of the standard is vehicle-to-roadside communication, but also some interaction with back-end systems (e.g., traffic management centers - TMCs) is included. 
No precise interaction and communication requirements are defined, but the communication is described in general terms. 
For the direct link, DSRC communication is assumed. 
Nevertheless, the objective is to define messages and data elements that are independent of the medium.

The aforementioned section stipulates that an RSU should engage in communication with the local traffic equipment and the TLC. 
However, it does not specify the manner in which this communication should occur or the type of information that should be transmitted. 

\textbf{Section 5.2.2} (Architecture) notes that \textit{authentication, encryption, privacy protection, misbehavior protection} \cite[p. 16]{TS19091} and other related topics are not within the scope of the standard. 
Furthermore, it is assumed that communication between all parties is \textit{secure}.
However, the term \textit{secure} is not defined in this context.

In \textbf{Section 5.2.3} (Message interaction), it is assumed that all messages will be be broadcast via DSRC and that no interactive message exchange will occur between two or more communication participants. 
It is further assumed that there is no reliability in the transmission of a message. 
Consequently, services must assume that messages can be lost and repeat the transmission several times until they receive a requested answer. 
Finally, concepts for enhanced spectral efficiency and reduced channel load, such as a zone of robust communication developed by \cite{ANH2013}, are not discussed. 
It is recommended that delivery should be assumed for services only if a broadcast from the recipient is received. 
However, it is not specified how many messages this should be. 
In the context of mobile vehicles, communication conditions change rapidly, so one message reception should not be assumed as sufficient to infer that a message was successfully delivered. 
Consequently, it is necessary to monitor the communication channel from a service site if, for example, a higher priority message is seen. 
Consequently, the aforementioned use case is no longer viable (e.g., the green wave for public transport is superseded by an emergency vehicle green wave, which is also known as the blue wave).

In \textbf{section 5.2.4} (Common operational assumptions), a list of fundamental communication assumptions is presented. 
\begin{itemize}
	\item It is assumed that all vehicles are equipped with a device that transmits CAM (or BSM) data, including information about location, speed, and direction. The minimum or maximum frequency at which this message is sent is not specified.
	\item It is further assumed that the communication range is up to 300 meters.
	\item It is recommended that RSU \textit{continuously} transmit SPaT messages and \textit{periodically} transmit MAP messages. With regards to MAP, the repetition interval should be based on \textit{RF propagation} and the \textit{speed of the approaching vehicles}. However, no further information is provided regarding what is meant by 'continuously' and which 'RF parameter' should be monitored.
	\item Priority vehicles require a priori knowledge of intersections to send the correct ID of SRM. The method by which they should obtain this information is not described.
	\item Prioritization at the RSU and the corresponding intersection will be based on the vehicle type (public transport, emergency vehicle, etc.).
	\item It is assumed that interaction between different RSU/TLC is necessary for platooning. This communication could be direct or via TMC.
	\item Specific security measures, that are expected and needed are not defined. For instance, is authentication sufficient, or should encryption also be possible? 
	\item It is of the utmost importance that messages should be media-independent. However, the protocol must be aware of bandwidth limitations and propagation latency. The manner in which a service acquires this information and responds is not defined. These characteristics are contingent upon the communication technology employed. Furthermore, additional parameters, such as packet size, jitter, and so forth, could be of interest. In the context of media-interdependent protocol design, this should be resolved at a lower layer and not by the service itself. 
\end{itemize}

\textbf{Section 5.3} ('Safety use cases') assumes DSRC communication, but the messages should be specified as technology agnostic. 
The use cases should assume that vehicles send CAM/BSM messages. 
Furthermore, it is stated that any protocol with the same characteristics as DSRC in latency, reliability, and range could be used. 
The same statement is given in sections 5.4 and 5.5 for the other use cases. 
It is notable that \textbf{section 5.2.3} only states an unreliable information exchange. 
Consequently, it is unclear what \textit{reliability} is expected.

\textbf{Section 5.3.3} states that \textit{real-time information} \cite[p. 18]{TS19091} will be transmitted. 
However, it is not clear what kind of \textbf{real-time} is meant, and it is unlikely that real-time communication will be expected. 

In \textbf{section 5.5}, a broadcast message flow is assumed, yet only two connection partners are specified. 
In the event that multiple vehicles would submit a request via an SRM, a broadcast would be necessary. 
Otherwise, a unicast could be employed. 
It is not assumed that a real-time connection exists between the TLC and the TMC. 
From a security perspective, it is assumed that the ID change (for vehicles to be pseudonymous) is off during a priority drive through an intersection and the corresponding message transfer. 
This raises the question of why a public transport vehicle or a public safety vehicle in action should change its ID. 
One could easily argue that such vehicles should not change their pseudonym because privacy is unnecessary. 
It is described that public safety vehicles need special authentication to claim priority rights. 
The European specifications that define the mechanisms for obtaining those rights (\cite{TS102940} and \cite{TS102942}) are not mentioned. 
Those rights are used for blue wave functions. 

Furthermore, it is assumed that public safety vehicles can transmit messages with a higher power on the channel. 
However, no reference for this is provided to substantiate this assumption. 
In Europe, the standards and norms \cite{ES202663}, \cite{EN302663}, \cite{EN303613}, and \cite{TS102724} for spectrum usage do not describe such a possibility. 
In the event that an emergency message must be transmitted, the aforementioned specification \cite[p. 11]{TS102724} outlines the possibility of doing so after a shorter guard interval on the control channel (CCH). 
However, the standard profiles from C2C-CC \cite{C2CCCBSP} and C-Roads \cite{CROADSHCP} do not indicate a similar approach. 

Finally, if vehicles were able to provide their route, the first RSU would be able to either inform the next RSU or contact the TMC to identify the traffic lights along this route.
This would enable the RSU to prepare the TLC for the anticipated arrival of priority vehicles, thereby allowing the TMC to preemptively prepare their signal program. 
However, it is not clear how this information should be communicated between the TLC and the TMC and vice versa. 
The SRM is linked to a specific intersection, which requires that priority vehicles have a list of intersections in advance. 

The functional description in \textbf{section 6} includes some vague requirements for \textit{performance measurement} \cite[p. 25]{TS19091}. 
The specifics of these requirements will be specified in the following standard version. 
Some of the requirements lack precision. 
Section 6.1.2 states that a public safety vehicle should only broadcast its information and priority if it is driving to or at the incident scene. 
However, what about the way an ambulance to the hospital? 
This is also a very critical situation. 
Two procedures are outlined for the signal pre-emption process to send SRMs. 
The first procedure is contingent upon certain conditions: the authorized vehicle is moving towards the intersection and requires a priority service, it has received at least one SPaT, and it has received a MAP or has obtained the intersection geometry in some other manner. 
The rationale behind this is not elucidated. 

The second procedure is contingent upon the use of a high-power OBU.
The first condition is the same as with the normal power OBU. 
The only other condition is the existence of the intersection geometry. 
The method of obtaining this geometry is not specified. 
The reception of a SPaT is not necessary. 
However, this approach has the disadvantage that a vehicle can send an SRM before receiving anything from the RSU at the intersection, but it will not be able to receive the answer SSM. 
Validation can be achieved via the SSM or SPaT messages if the request was received and granted.
However, this is only possible if the vehicle is within the receiving range. 
This is because the transmit power is only effective in one direction, so an increase in transmit power on one side will only affect that side. 
In contrast, a higher antenna gain or a more elevated location would be effective in both directions. 
In the case of high-power transmission, there is the potential for a reduction in the processing time on the RSU and the corresponding TLC. 
The public safety vehicle would be required to send the request multiple times due to the uncertainty surrounding the receipt of the request by the RSU. 
Additionally, the RSU would be required to send the SSM with greater frequency. 
The sending interval of the SPaT should remain unaffected. 
This procedure is not spectrally efficient, and it is unclear how much time gain can be gained in comparison to other critical information that may be delayed in the vicinity of the high-power vehicle due to this. 
Further research is necessary to investigate the implications.

In the aforementioned scenario, it is of utmost importance that the SSM, the public safety vehicle, and the intersection be identified with a unique ID. 
In addition, the vehicle ID should be supplemented with a vehicle class (public transport, commercial, military, ambulance, police vehicle, etc.). 
This classification will enable the TLC to prioritize vehicle requests based on their respective class. 
Moreover, the vehicle should include the time it will be at the intersection, its current position, speed, and heading, as well as the period during which the request should be valid. 
It is imperative that the vehicle ID remains constant throughout the pre-emption process, as it serves as a unique identifier for tracking the vehicles by the RSU. 
The standard procedure for ID change is outlined in chapter \ref{Chapter_ITSProtocols-TP-ITSApp-secp}.

The same assumptions and conditions are applicable to public transport and commercial vehicles, with the exception of the high-power case, which is unavailable. 
Furthermore, a public service vehicle should include its service status, which could be indicated by open doors or by passenger load information. 
The latter could be used to prioritize public transport vehicles with more passengers on board. 
Additionally, the SRM should include information about the schedule of the public transport vehicle and its punctuality.

Information regarding the road geometry (intersections, speed curves, specific locations, etc.) can be transmitted by an RSU. 
While this is not explicitly stated, it is also possible for a public safety vehicle to disseminate such information to connected and automated vehicles in order to inform them of an incident and the mandatory deviations from the regular road geometry.
The geometry encompasses a reference point, lanes, approaches, coordinates for the street lanes and approaches, permitted vehicle maneuvers, and road and crossing information for pedestrians and cyclists. 
Furthermore, information regarding other signs, gates, flashing beacons, and information pertaining to signal groups can be included. 
Additionally, information about the pavement should be transmitted to connected vehicles. 
It is not specified which message type should be utilized. 
It can be assumed that the MAP message should be facilitated.

In addition, the RSU should disseminate GNSS augmentation information. Two distinct message types have been defined for this purpose.
\begin{itemize}
  \item National Marine Electronics Association (NMEA) 0183 \cite{NMEA0183} differential GPS correction messages. (current version: NMEA 2000)
  \item Radio Technical Commission for Maritime Services (RTCM) 10402.3 and 10403.1 messages (current version: 10403.3 from 2016 \cite{RTCM104033}). This message is capable of functioning with the NAVSTAR-GPS, as well as GLONASS, Galileo, BDS, and QZSS.
\end{itemize} 

No information is provided regarding the type of message to be utilized. 
It can be reasonably assumed that the RTCM corrections message will be employed.

An intersection is required to transmit signal phase and timing information (SPaT) messages via a TLC and RSU. 
These messages must be broadcast in a timely manner, although no further information is provided regarding the definition of a timely manner. 
The message should include an ID of the intersection, an ID to the corresponding MAP message, signal group information, allowed maneuvers, intersection status, pedestrian status, special state, time stamps, the succeeding maneuvers, and signal indication time, detected pedestrians, pedestrian call, optimal speed per lane, lane storage availability, and possible wait indications.

Information regarding cross-traffic and detected VRUs should be transmitted via an RSU at intersections. 
No specifications are provided regarding the message types to be utilized. 
The transmission of information regarding vehicles that have violated traffic rules (e.g., entering the intersection after the signal had turned red) is also subject to the same conditions.

It is recommended that the RSU transmits SSM the current status of the SRM. 
No interval for repetition has been specified, and the timeframe for sending the message is unclear. 
It is assumed that the message will be sent until the TLC attempts to serve the request. 
The SSM should include the information on what request is currently being processed, the current status of the intersection, the request, and the requesting vehicle (the ID used in the request).

In order to enhance the system performance and reduce the bandwidth, it is recommended that the RSU reference only one global reference point in the MAP, and that all lanes be referenced only by delta coordinates.

It is recommended that the SRM be transmitted with a maximum frequency of 2 Hz. 
The default maximum response time for the RSU/TLC should be 0.5 Hz. 
Subsequently, the SSM should be transmitted with a minimum frequency of 2 Hz. 
The RSU should transmit the SSM until the requesting vehicle has cleared the intersection. 
The stipulation set forth in chapter 6.14.4 is in conflict with the stipulation in chapter 6.11, which states that the SSM should be sent only until the TLC has adjusted the signal plan to accommodate the request. 
The situation is not clear, and it is possible that this may lead to misinterpretations. 

With regard to the roadway geometric information, it is stated that an RSU transmitting such messages is a safety-critical RSU, given that the messages are safety-related. 
Furthermore, the fact that wireless communication is \textit{not 100\% reliable} \cite[p. 42]{TS19091} is cited as the rationale for the differing message transmission rate. 
However, the rationale behind the differing transmission rates is not entirely clear.
However, a transmission rate between 0.5 Hz and 2 Hz is specified with a default value of 1 Hz. 

It is recommended that the GNSS augmentation-related messages be transmitted at a minimum frequency of 1 Hz, while any auxiliary messages should be sent at least every 15 seconds. 
However, no specifications are provided regarding the nature of the aforementioned auxiliary messages.

The transmission rate of SPaT messages is contingent upon the considerations of roadway geometric information messages. 
The optimal transmission interval is between 0.5 Hz and 10 Hz, with a default interval of 6.$\overline{6}$ Hz (every 150 ms).

In terms of the broadcast of VRU sensor information, the same considerations are made as for SPaT and roadway geometric information. 
The transmission intervals are identical to those for SPaT.

It is noteworthy that the reliability of the communication channel is a factor in the case of SRM and SSM. 
This is not explicitly stated, as is the case with all other messages.

\textbf{Sections 7} ('Message') and \textbf{8} ('Conformance') do not contain any information pertaining to communication.

\textbf{Annex A} specifies certain communication-related requirements: 
\begin{itemize}
	\item \textbf{Communication systems:}  RSU, OBU (vehicle), TLC, and TMC
	\item \textbf{Communication technologies:} DSRC-V2I, wide-area broadband/network communication vehicle-2-TMC (e.g. \cite[p. 89]{TS19091}), RSU-2-RSU communication (e.g. \cite[p. 54]{TS19091}), RSU-2-TMC communication (e.g. \cite[p. 72]{TS19091}), and VRU-2-RSU communication (e.g. \cite[p. 88]{TS19091})
	\item \textbf{Communication roles:} data transmitter, data receiver
	\item \textbf{Messages:} (messages only related to traffic management are omitted in the following list)
		\begin{itemize}
			\item SPaT, including an indication of VRU presence 
			\item MAP includes information on inductive charging. Although this is not currently reflected in the format, it is expected to be included in the future.
			\item BSM or CAM
			\item SRM 
			\item SSM
			\item TIM
			\item A message, the content of which is not specified, is transmitted to vehicles in order to inform them of the presence of conflicting traffic at intersections.
			\item RTCM corrections
			\item \textit{Sensor information} (e.g. \cite[p. 84ff]{TS19091}) about conflicting vehicles and VRU be transmitted. This message should include the detected objects' coverage, location, speed, heading, and acceleration. It has been proposed that the definition of such a message is an open issue. 
			\item The transfer of presence information from VRU to a RSU may include the trajectory and speed of the VRU (see \cite[p. 89]{TS19091})
			\item A message that is not specified for the purpose of 'measuring performance data' (e.g., \cite[p. 93]{TS19091}).
			\item A lack of clarity regarding the intended message for speed limit or speed corridor information (e.g., \cite[p. 101]{TS19091}) is identified as a potential issue.
		\end{itemize}
	\item \textbf{Communication constraints:} 
		\begin{itemize}
			\item In addition to DSRC, \textit{other medium[s] that meet the performance requirements for the use case} (\cite[p. 47]{TS19091}) are also permitted. However, no performance requirements are given beyond those pertaining to transmission rates. It should be noted that certain use cases, such as PR4, do not permit the use of other technologies in lieu of DSRC. 
			\item Additionally, no example is provided for \textit{wide-area broadband communications} (e.g., \cite[p. 47]{TS19091}). Furthermore, no rationale is presented for the necessity of broadband communication and the potential for narrow-band communication.
			\item The communication between the RSU and RSU is not adequately explained. In certain use cases, the preconditions, flow (main and alternate), and issues do not align with the communication in question (e.g., use case PR2 \cite[pp. 53ff]{TS19091}).
			\item In some instances, the initiation of a use case is contingent upon the vehicle being within the \textit{radio range} of an RSU.
			\item In certain instances, the communication between a vehicle and the TMC is depicted via a WAN (e.g., \cite[p. 58]{TS19091}). In other cases, this is absent (e.g., \cite[p. 50]{TS19091}), resulting in the exchange of identical messages.
			\item In the use case PR3-a \cite[p. 62]{TS19091}, it is necessary for an RSU to ascertain the size of a platoon. However, the methods of obtaining or transmitting this information are neither specified nor explained.
			\item It is recommended that the RSU be able to send \textit{vehicles served performance measures} \cite[p. 69]{TS19091}, but there is no indication provided regarding the specific type of performance being measured or the method of communication employed.
			\item The SSM does not include a field for recording instances of conflicting vehicle requests. For instance, two ambulances may traverse different roads to reach the same intersection, where they can communicate with the RSU at the crossing. However, they may be unable to reach each other. In such a case, the RSU should communicate this situation \cite[p. 70]{TS19091}. Nevertheless, the document only specifies that this could be tracked by a TMC (it is unclear why not an RSU) and communicated via an RSU or \textit{other wide-area wireless service/media} \cite[p. 70]{TS19091}. However, it is not specified what kind of message could be used. The European profile (Annex G) introduced a reason for a not served preemption.
			\item In the event of an emergency, the TMC is responsible for establishing a blue wave-green wave. The method of communication with the local TLC or RSU is not specified.
			\item The previously described and explained mechanism of high-power OBU systems is employed once more. In this instance, it is assumed that the topology and route are already known.
			\item The \textit{Vehicles' DSRC transmit performance under high-speed approach conditions} should be \textit{adequate for timely RSU reception} \cite[p. 78]{TS19091}. However, it is not clear what is meant by the term 'high-speed' or what is meant by the term 'timely'.
			\item The \textit{Vehicles' DSRC transmit performance under congested conditions} should be \textit{adequate for timely RSU reception} \cite[p. 95]{TS19091}. The precise meaning of the terms 'congested' and 'timely' is not defined.
			\item An RSU should be capable of acquiring and broadcasting information regarding the condition of the road surface, including weather conditions. Currently, only BSM is able to perform this function, although CAM is not.
			\item MAP and SPaT should be transmitted in \textit{real-time}, as indicated in \cite[p. 80]{TS19091}. It is not within the purview of the communication system to stipulate the specific type of \textit{real-time} implementation.
		\end{itemize}
	\item \textbf{Other constraints:}
		\begin{itemize}
			\item Vehicle's exact positioning within a lane. 
			\item Pre-defined routes within the vehicle.
			\item An infinite platoon is assumed \cite[p. 60]{TS19091}
			\item A security management system should allow an OBU to check the message coming from an RSU (e.g. \cite[p. 81]{TS19091}). In this case, messages should be authentic. 
		\end{itemize}
\end{itemize}

An illustration for each use case should provide a clear description of the use case in question. 
While the diagrams include some numbering for the traffic flow, the numbers are neither explained in the diagram nor referenced in the accompanying description. 

\textbf{Annex B} provides a list of message elements for SPaT, MAP, SRM, and SSM. In the case of SRM and SSM, a VehicleID is included, which is used in conjunction with speed and position to identify the vehicle requesting the signal and which request is served. 
Furthermore, SPaT includes a Boolean value for identification in the event that conflicting VRU are present in relevant zones at an intersection.

The final \textbf{Annex G} contains the regional extensions of the \cite{SAEJ2735} for Europe. 
A list of ITS-capable station positions is added in the field \textit{ConnectionManeuverAssist-addGrpC}. 
This includes the identification, exact position (referencing the high-precision reference position of the MAP), and time reference of the message from which the information was derived.  

In conclusion, \cite{TS19091} specifies the usage of SPaT, MAP, SSM, and SRM based on the definition of \cite{SAEJ2735}. 
While the RTCM message is mentioned, it is not as thoroughly addressed as the four other messages.
Additionally, several other messages (\textit{performance}, \textit{auxiliary GNSS information}, \textit{sensor information}, etc.) are mentioned, but neither definition is provided nor a link to other standards is provided.

With regard to communication, the standard endeavors to motivate the requirements through the use of several use cases in three categories (safety, mobility, and priority). 
However, the communication-related requirements are often unclear, and contradictions arise in different standard sections. 
Therefore, a review phase would be advisable to clarify these issues.

There are precise requirements regarding sending rates for all four messages. 
However, the use of terms such as 'timely,' 'real-time,' 'high-speed,' and 'radio range' without accompanying explanations makes it difficult to ascertain the exact nature of the communication system's requirements.

The specification identifies several communication relationships, including Vehicle-2-RSU, Vehicle-2-TMC, RSU-2-RSU, RSU-2-TMC, and VRU-2-RSU.
However, only Vehicle-2-RSU is associated with a technology, namely DSRC, or an equivalent technology for similar use cases. 
To communicate with the TMC, a \textit{wide area broadband/network communication} should be employed. 
However, the specific requirements for this technology and the necessity of broadband communication remain unclear. 

Some effort has been made to reduce the size of the message by referencing values in other messages or cross-link information (e.g., delta-positions). 
Additionally, message fragmentation has been considered (only for MAP) but only for separating a message that would otherwise be too big. 
No indication is provided regarding intelligent fragmentation, where information from one message would be sufficient for the majority of use cases and more detail could be sent less frequently, thus conserving bandwidth and optimizing spectrum use. 
The standard does not include any suggestions for intelligent message distribution based on the number of vehicles or the presence of any particular vehicle. 
Concepts such as a \textit{zone of robust communication} \cite{ANH2013} could be a useful starting point for more efficient communication. 
Additionally, the transmission of SPaT and MAP messages in the absence of a nearby vehicle to receive them is unnecessary. 
The transmission of these messages could, for instance, commence upon the receipt of a BSM or CAM from a vehicle. 
This represents an open research topic regarding the optimal methodology for serving all use cases (in particular those related to safety), while simultaneously reducing the number of messages sent.

It is notable that the security aspect of the standard is absent. 
It is stated that security is assumed and that vehicles and the RSU should be able to verify the authenticity of a message. 
However, no further detail or references are provided. 
Regarding privacy, it is indicated that vehicles that are sending SRM to obtain a preemption at an intersection should not alter their ID throughout the process. 
The RSU transmits sensor data (which may include communication, camera, LIDAR, or other information) about detected vehicles or VRU, including IDs, positions, speed, and heading. 
This data is also stored by the RSU to provide an overview of all vehicles and RSU in and around the intersection and to warn traffic participants of potentially critical situations, thereby preventing accidents. 

\subsection{TPEG2 - ISO 21219}
\label{Appendix_PE-TP-TPEG2}
The ISO 21219 standard family is comprised of the following parts, as of 2024:
\begin{itemize}
	\item \textbf{Part 1} Introduction, numbering, and versions (TPEG2-INV) (deprecated)
	\item \textbf{Part 2} UML modeling rules (TPEG2-UMR)
	\item \textbf{Part 3} UML to binary conversion rules (TPEG2-UBCR)
	\item \textbf{Part 4} UML to XML conversion rules
	\item \textbf{Part 5} Service framework (TPEG2-SFW)
	\item \textbf{Part 6} Message management container (TPEG2-MMC)
	\item \textbf{Part 7} Location referencing container (TPEG2-LRC)
	\item \textbf{Part 8} reserved for future use
	\item \textbf{Part 9} Service and network information (TPEG2-SNI)
	\item \textbf{Part 10} Conditional access information (TPEG2-CAI)
	\item \textbf{Part 11} reserved for future use
	\item \textbf{Part 12} reserved for future use
	\item \textbf{Part 13} reserved for future use
	\item \textbf{Part 14} Parking information application (TPEG2-PKI)
	\item \textbf{Part 15} Traffic event compact (TPEG2-TEC)
	\item \textbf{Part 16} Fuel price information and availability (TPEG2-FPI)
	\item \textbf{Part 17} Speed information (TPEG2-SPI)
	\item \textbf{Part 18} Traffic flow and prediction application (TPEG2-TFP)
	\item \textbf{Part 19} Weather information (TPEG2-WEA)
	\item \textbf{Part 20} reserved for future use
	\item \textbf{Part 21} Geographical location referencing(TPEG-GLR)
	\item \textbf{Part 22} OpenLR\footnote{OpenLR is an open specification for map agnostic location information developed by TomTom; OpenLR homepage: \url{https://www.openlr-association.com/}} location referencing (TPEG2-OLR)
	\item \textbf{Part 23} Road and multimodal routes (TPEG2-RMR)
	\item \textbf{Part 24} Light encryption (TPEG2-LTE) 
	\item \textbf{Part 25} Electromobility charging information (TPEG2-EMI) 
	\item \textbf{Part 26} Vigilance location information (TPEG2-VLI)
\end{itemize}

For the purpose of this evaluation, only certain aspects of the standard are of interest. 
As an illustration, part 15 is described in detail.

The ISO TS 21219-15 Traffic Event Compact (TPEG-TEC) \cite{TS21219-15} provides information about traffic events such as road works, black ice, and traffic jams. 
The latter two fall within the category of high-priority local hazard warnings. 
These are analogous to the ITS DENM messages (see chapter \ref{Chapter_ITSProtocols-TP-ITSApp-DENM}) for the traffic protocol sector. 
In contrast, the information provided regarding road works may include detailed data about the specific road works in question.
This may include the number of lanes affected, the speed limits in place, and the estimated time loss.
Additionally, is may include information about potential detour routes, which may be restricted to certain vehicle types.
Such information could include details about which streets should be avoided when calculating alternative routes, such as those that pass by a kindergarten.

\begin{enumerate}
	\item[1-4] The initial four chapters delineate the \textbf{Scope}, \textbf{Normative references}, \textbf{Terms and definitions}, and \textbf{Abbreviated terms}. 
	\item[5] \textbf{Application-specific constraints} This chapter describes the structure of the TPEG2 message, the usage of the overall message management container, and possible extensions for TEC. 
	\item[6] \textbf{TEC structure} The UML class diagram represents the structure of the TEC message. 
	\item[7] \textbf{TEC message components} The message classes presented in chapter 6 are detailed in this chapter. The TEC message comprises of three components: a mandatory management container, one or none event descriptions, and one or none location referencing containers. An event is defined as a discrete occurrence accompanied by the following information: the nature of the occurrence, the impact of the occurrence, the location of the occurrence, the restrictions (e.g., speed, lanes, roads, etc.), the temporal boundaries of the occurrence, and potential alternative routes.
	\item[8] \textbf{TEC Datatypes} The data types, as depicted in the UML diagram designated 'DataStructure,' represent a specific class of objects that must be unique. Here, the speed limit for a specific road segment, the nature of restriction for a particular road, and the segment of an alternative route are described.
	\item[9] \textbf{TEC Tables} The chapter describes the potential values that each class attribute may assume.
	\item[A] \textbf{Annex A} The annex provides a detailed account of the binary representation rules that apply to the TEC message.
	\item[B] \textbf{Annex B} The annex provides a detailed account of the XML representation rules that apply to the TEC message.
\end{enumerate}

In conclusion, the TEC standard from TPEG describes only the message itself, without including any communication-relevant parameters.

\subsection{Vehicle-2-Grid Communication}
\label{Appendix_PE-TP-V2G}

This chapter outlines the specifications regarding the communication requirements (ISO 15118-1) and the second-generation protocol (ISO 15118-20).

\subsubsection{ISO 15118-1}

The ISO 15118-1 \cite{ISO15118-1} standard delineates the information, use cases, and requirements for V2G communication. 
It is contingent upon \cite{TS101556-3} (as detailed in chapter \ref{Chapter_ITSProtocols-TP-ITSApp-EV}) as a potential communication application protocol. 
Additionally, it depends on the IEC 61851\footnote{IEC 61851: Electric vehicle conductive charging system \url{https://webstore.iec.ch/publication/33644}} standard for electric vehicle conductive charging systems and the IEC 61980\footnote{IEC 61980: Electric vehicle wireless power transfer (WPT) systems \url{https://webstore.iec.ch/publication/22951}} standard for electric vehicle wireless power transfer (WPT) systems.

\begin{enumerate}
	\item[1-4] The initial four chapters delineate the \textbf{Scope}, \textbf{Normative references}, \textbf{Terms and definitions}, and \textbf{Abbreviated terms}. 
	\item[5] \textbf{Requirements} The requirements are structured in the following parts: general communication requirements, user-specific requirements, OEM-specific requirements, utility-specific requirements, wireless communication requirements, RPT requirements, and traceability requirements. The actual requirements are listed in Annex E of the standard. 
	\item[6] \textbf{Actors} ISO 15118 identifies two categories of actors: primary and secondary. The primary category encompasses the electric vehicle (EV) and its components, as well as the electric vehicle supply equipment and its components. ISO 15118 describes the interface between the vehicle and the power supply equipment. The secondary category, for the sake of completeness, includes entities such as e-mobility service providers, clearinghouses, charging station operators, fleet operators, electricity providers, and so forth. However, these entities are not described in detail.
	\item[7] \textbf{Use case elements} The use case elements are structured in ten distinct functions (compare \cite[p. 25]{ISO15118-1}):
		\begin{enumerate}[label=(\Alph*)]
			\item Start of the communication session
			\item Communication set-up
			\item Certificate handling
			\item Identification, authentication, and authorization
			\item[(P)] Paring and fine positioning
			\item Target setting and energy transfer scheduling
			\item Energy transfer controlling and re-scheduling
			\item Value-added services
			\item End of the energy transfer period
			\item ACD connect/disconnect
		\end{enumerate}
		Subsequently, the task groups are comprised of several use cases, commencing with the letter of the task group, or, in the case of wireless communication, with an additional \textit{W} preceding the letter of the task group.
	\item[A] \textbf{Annex A} The 'conductive charging infrastructure architecture' delineates the systems engaged in communication and energy transfer, the distinct network characteristics, and the variations of the system architecture. Furthermore, it introduces timing and loading patterns and additional equipment for energy scheduling.
	\item[B] \textbf{Annex B} The 'security' annex presents the standard cryptographic and privacy goals, including 'confidentiality,' 'data integrity,' 'authentication,' 'non-repudiation,' 'privacy,' and as an additional goal 'reliability/availability.'
	\item[C] \textbf{Annex C} In the initial sections of the annex, the use case elements and scenarios are outlined, illustrating how they can be combined. The subsequent sections address the wireless communication scenarios and their distinct characteristics, particularly the various charging possibilities, including wireless, cable, and ACD. Finally, the annex presents a few general assumptions regarding electric vehicle charging.
	\item[D] \textbf{Annex D} This annex describes the variants for a 'typical RPT system' for forward and reverse power transfer to and from the vehicle.
	\item[E] \textbf{Annex E} The normative 'requirements list,' comprising 94 requirements, is presented in the final annex and provides a summary of the description presented in chapter 6 of the standard.
\end{enumerate}

The initial four sections of the standard \cite{ISO15118-1} do not contain any relevant information for the present analysis. 

There are minor discrepancies between the Requirements chapter and Annex E, which are likely due to editorial errors. 
For instance, the requirement regarding instances when customer-specific information can be modified in an electric vehicle is absent from Annex E, despite the fact that a vehicle sold as a pre-owned automobile is also subject to this requirement. 

Two types of communication are mentioned: basic signaling and high-level communication.
However, only the letter is specified in ISO 15118-1. 
No definition is given for \textit{basic signaling}. 
Additionally, the document only defines requirements and does not specify protocols or data for the transmission. 
Communication is only considered between an EV and the charging infrastructure and between the charging infrastructure and a secondary actor (only the first one is specified). 
The communication between an EV and a secondary actor is not addresses in the document. 
From a communication perspective, three general goals are listed: ISO 15118-3 or ISO 15118-8 should be used between the EV and the charging infrastructure, all data between the EV and the secondary actors should be confidential, and that all communication should be protected against \textit{modification and imitation (hacking)} \cite[p. 16]{ISO15118-1}. 
However, there is no mention of the confidentiality of the communication between the EV and the charging infrastructure.

The privacy requirements can be interpreted as being very strict. 
Private information shall only be readable for the intended recipient and should only be transferred if it is otherwise impossible to fulfill the user's request. 
Additionally, the user should be informed of which individuals have access to which parts of their private data.

In 'Annex B - Security', the necessity of authentication is indicated in certain instances, yet no explicit requirement exists for that purpose. 
With regard to the communication setup, authentication is indicated solely for wireless communication. 
In contrast, no authentication is required for wired communication. 
Authorization is accomplished via identification numbers. 

In the context of wireless communication, no explicit technology is specified. 
However, the charging infrastructure should broadcast the necessary information for the EV, such as discovery.

With regard to the matter of trivia, it is indicated that 'high-level communication' is employed exclusively in the event that both the EV and the \textit{supply equipment} are equipped with a communication device that is capable of 'high-level communication.' 

In the use case task groups \cite[pp. 25ff]{ISO15118-1}, the communication set-up phase is described. 
It is recommended that the phases should contain a mechanism to exchange information about the capabilities of the communicating parties, such as protocol standards and versions. 
The described certificate installation process is mentioned as one possibility, but from a security point of view, it should not be used. 
In the process, the EV requests a certificate installation. 
The request is transmitted to the charging infrastructure, which then forwards it to the secondary actor responsible for providing the certificate. 
This secondary actor returns the certificate and the corresponding private key. 
At the very least, the latter should be encrypted with the certificate of the EV. 
The \textit{Identification and authorization} process, as described in \cite[pp. 35ff]{ISO15118-1}, is only superficially authenticated, as both the EV and the charging infrastructure transmit their IDs (e-mobility authentication identifier and EV supply equipment identifier). 
The EVID or a contract certificate may be forwarded to a secondary actor to authorize the charging process. 
If the authorization process is successful, the vehicle and the infrastructure negotiate the charging details and constantly update each other during the charging process.

'Annex A' presents a number of potential architectural solutions. 
The communication and relevant players are illustrated using the ISO/OSI seven-layer model. 
Given that the charging infrastructure frequently comprises a single charging point, variations between one-to-one and one-to-many communication scenarios between the charging infrastructure and the EV are presented. 
Furthermore, the timing and scheduling of the power transfers are critical factors and are described in detail. 

As previously stated, 'Annex B - Security' outlines six security objectives for literature derived from the 1996 publication \textit{Handbook of Applied Cryptography} by Menezes, van Oorschot, and Vanstone. 
These objectives should be aligned with contemporary advancements in security and privacy. 
Table \ref{table_Appendix_PE-TP-V2G} compares the ISO 15118 goals to the GDPR\footnote{GDPR website: \url{https://eur-lex.europa.eu/legal-content/EN/TXT/?uri=CELEX\%3A02016R0679-20160504}}/ SDM\footnote{SDM website: \url{https://www.datenschutzzentrum.de/uploads/sdm/SDM-Methodology_V2.0b.pdf}} objectives. This comparison is not exhaustive, but it indicates whether the development of the standard aligns with the current European rules for security and data protection.

\begin{table*}
    \caption[]{Comparison of ISO 15118 and GDPR/SDM security goals} 
    \label{table_Appendix_PE-TP-V2G}
    \centering
    \thispagestyle{empty}
    \begin{tabular}{p{3cm}p{3cm}p{10cm}}
	\toprule
	\textbf{ISO 15118} & \textbf{GDPR} & \textbf{Comment} \\ 
	\midrule
		Confidentiality & Confidentiality & ISO 15118 offers a more expansive definition than that of the GDPR. While the GDPR restricts access to data subjects, ISO 15118 limits confidentiality to all authorized parties. \\ \hline
		Data Integrity & Integrity & In terms of data integrity, both standards and regulations define the same concept. The GDPR definition encompasses a broader range of meanings than the standard definition. In addition to defining data as a resource that should be used for the intended purpose only and kept up to date, the GDPR also requires that data subjects be informed of the processing of their data. \\ \hline
		Authentication & Intervenability & The ISO 15118 standard delinates the authentication of data (source, date, time, etc.) and communication endpoints. However, the necessity of mutual authentication is not explicitly stated in the standard. Furthermore, it is unclear who should be authenticated, whether it should be a legitimate participant or a distinct person or system. The GDPR defines the authentication process in the context of authenticating the data subject by a data controller. This process is designed to ensure that the subject is the one who they claim to be in order to access their data. \\ \hline
		Non-repudiation & / & The objective is to ensure the integrity and authenticity of the data. When these two goals are met, the originator is unable to deny the data.  \\ \hline
		Privacy & Data minimization, Unlinkability, Transparency, and Intervenability & ISO 15118 defines privacy as \textit{related to personal data protection}, \cite[p.82]{ISO15118-1}, and subsequently provides a vague definition of personal data. In contrast, the GDPR encompasses a broad spectrum of privacy-related goals. The fundamental requirement for data protection is data minimization. This is followed by unlinkability, which stipulates that personal data may not be linked with any other data for reasons other than those originally agreed upon. Transparency is also a key aspect of data protection. It requires that a data subject be able to ascertain precisely what data is collected for what purpose and by whom. Finally, intervenability is also a crucial aspect for data protection. It allows data subjects to inform, correct, restrict, delete, migrate, and withdraw their consent for data processing.  \\ \hline
		Reliability / Availability & Availability & In accordance with the ISO 15118 standard, a service should be available and communication should be reliable. In terms of the GDPR, data should be retrievable and presentable to the human user at any time. \\ \hline
    \bottomrule
    \end{tabular}
\end{table*}

In conclusion, in \cite{ISO15118-1}, no explicit security requirements are made for the connection between the EV and the charging infrastructure, despite the fact that private information (e.g., for billing) is exchanged and that private information should \textit{only be readable by the intended recipient} \cite[p. 17]{ISO15118-1}.

It is inadvisable to permit the secondary actor to generate the private key. 
The private key should be stored exclusively within the EV in a secure enclave and should never leave the EV. 
A more optimal approach would be for the EV to generate a certificate signing request, and for the secondary actor to sign the request (as is commonly done for certificates associated with websites).

The standard makes reference to the ETSI TS 101 556-3 \cite{TS101556-3} communication specification in the bibliography. 
Regrettably, only one of the twenty-four references is linked by ID in the text. 
The ETSI standard is mentioned in the wireless use case 'discovery with reservation', where it is stated that the reservation ID can be obtained via \cite{TS101556-3}.

\subsubsection{ISO 15118-20}
ISO 15118-20 \cite{ISO15118-20} delineates the requirements for the second-generation network and application protocol for vehicle-to-grid communication.
With regard to the OSI reference model, the initial two layers (the physical layer and the data link layer) are described in ISO 15118-3 and ISO 15118-4, respectively.
All other communication-related specifications are outlined in this document.
with regard to the network layer, IPv6 is employed.
The document makes no mention of IPv4, and therefore it should not be used.
The layer is referred to as the 'network link layer,' which is somewhat confusing.
The two typical transport layer protocols (UDP and TCP) are possible, and no indication is given about preference.
Transport Layer Security (TLS), preferably in version 1.3, is used on the session layer.
The presentation layer defines the data encoding based on XML to binary, designated as Efficient XML Interchange (EXI), as specified by W3C.
The transport protocol is specified to be transferred via TCP.
This protocol is designated as the Vehicle to Grid transfer protocol (V2GTP).
it is notable that UDP is not mentioned as a possibility.

A vehicle willing to charge must undergo different procedures depending on the type of connection (cable or wireless).
The cable-based connection is described in the 15118-{3,4} specifications.
For WLAN, the \cite{ISO15118-20} provides some guidance.
It does not provide information about SSID naming to be used or mandate security mechanisms.
The use of IEEE 802.1X for authentication and authorization is recommended. 
IEEE 802.1X is a widely utilized protocol in wireless networks, despite its designation as a wired interface-specific standard.
The authentication process involves a mutual authentication via a remote authentication dial-in user service (RADIUS), where the client verifies the server's certificate and the server verifies the client's certificate.
To enhance security, it is recommended to deploy WPA3-Enterprise or WPA2-Enterprise mechanisms.

A vehicle must utilize the Supply Equipment Communication Controller Discovery Protocol (SDP) to identify its charging station via the network.
A discovery message is transmitted by a vehicle to all nodes on the local-link multicast address FF02:1 (as specified in RFC 4291) using UDP with port 15118.
All charging stations on the network on the link can then respond and provide their IP address and port.
The payload includes the security and transport protocol information for the data exchange.
Currently, only TLS and TCP are specified.
The communication via SDP, which is V2GTP, is made unsecured.
This emphasizes the need for security on the lower layer and why unencrypted wireless LAN connections should be avoided.
In a WLAN setup, additional information is exchanged (e.g., SSID).

The V2GTP is a lightweight protocol with a fixed eight-byte header and a payload of up to four gigabytes.
The header includes the protocol version, the inverse protocol version, the payload type, and the payload length.
It is noteworthy that the designers have chosen to commence the numbering at the hexadecimal value of $0x8000$, which is precisely a factor of ten to the EtherType\footnote{List of EtherTypes is available on the IANA webpage: \url{https://www.iana.org/assignments/ieee-802-numbers/ieee-802-numbers.xhtml}} of IPv4 ($0x0800$) as used for Ethernet \cite{IEEE802.3}.
Additionally, the EtherType does not utilize the values from $0x0000$ to $0x05DC$, as they are employed for the IEEE 802.3 length field.
The rationale behind the decision to leave the initial $2^{26}$ values unassigned is not elucidated.

The specification places significant emphasis on the use of security and a variety of different types of certificates.
It is recommended that certificates of different types be employed for the purpose of network authentication, encryption, authorization, and contracts.
The use of a Trusted Platform Module (TMP) in version 2 is strongly recommended for the storage of private keys.
The different trust relations and the certificate exchange are described in detail.
The authentication and encryption via TLS are based on mutual authentication and the usage of the Online Certificate Status Protocol (OCSP), which allows for the verification of the validity of a certificate or the revocation thereof.
Additional security mechanisms based on the W3C recommendation for XML signatures are described at the application layer.
\cite{ISO15118-20} even describes firewall rules with IP addresses and ports to be cleared.
In this context, it is unclear whether port 1080 should be available.
While it is not explicitly stated elsewhere in the document, the purpose of the port and the circumstances under which it should be used remain unclear.

The application layer messages encompass a handshake message, a message for maintenance, request and response messages for distinct use cases, configuration messages, error handling, multiplexing, and synchronization procedures.
All messages are defined in XML.


\subsection{Service Announcement}
\label{Appendix_PE-CP-ITSCom-SA}

The Service Announcement Message (SAM) (\cite{TS102890-1}, \cite{EN302890-1}) is a media-independent protocol for the distribution of a list of services that a station provides for a specific geographical area. 
Any station can provide traffic-related services. 
The definitions of framework conditions and the SAM are the central part of the standard. 
It is important to note that the ETSI \cite{EN302890-1} is based on ISO \cite{TS16460} and only defines a specific profile and extension of the ISO standard. 
The SAM message specifies the type of service provided, the access technology, and the channel where the service can be consumed. 
This must not be the communication technology or channel where the SAM is transmitted. 
In most cases, it will not be the same channel because the receiving station already receives the service messages.

In the ASN.1 description\footnote{ETSI ASN.1 repository: \url{https://forge.etsi.org/rep/ITS/asn1}} of \cite{TS16460}, the following communication technologies are possible, based on different ISO ITS/CALM profiles:
\begin{itemize}
	\item unknown
	\item any
	\item Cellular (explicitly 2G, 3G, LTE)
	\item Infra-red
	\item Microwave in 5 GHz (based on \cite{IEEE802.11})
	\item Millimeter-wave from 57 GHz to 66 GHz
	\item WiMAX (based on \cite{IEEE802.16})
	\item MWB below 6 GHz (HC-SDMA)
	\item MBWA (based on \cite{IEEE802.20}, the standard was withdrawn in 2008 and the amendment in 2010)
	\item 6LoWPAN (based on \cite{IEEE802.15.4} and IETF RFCs 4944, 4919, and 6282)
	\item DSRC application layer (based on \cite{ISO15628})
	\item CAN (based on ISO 11898-\{1-6\} and SAE J2284-\{1-5\})
	\item Ethernet (based on \cite{IEEE802.3})
\end{itemize}

In the future, new and enhanced technologies such as C-V2X, DAB+, RFID, or Bluetooth should be supported. 
The specification indicates that these enhancements must be backward compatible at the same time.
However, the specifications of some other fields are not that precise. 
For instance, the SAM should be sent as a broadcast with a repetition interval between 0 and 255. 
It is unclear if this is an interval or a frequency because no unit is provided. 
It is conceivable that the precise definition as outlined in \cite{IEEE1609-3} will be employed. 
In that document, the value represents the number of times the SAM is transmitted per five seconds. 
A similar approach is taken with the data rate DataRate80211 and transmit power TXpower80211, where only integer values are permitted between 0 and 255 or between -128 and 127, respectively. 
No additional information about the unit is provided.  
According to \cite{IEEE1609-3}, the value of the DataRate80211 could be data rates between 1 Mb/s and 63.5 Mb/s, with a resolution of 500 kbit/s. 
The value should be interpreted as the minimum data rate required. 
Furthermore, based on \cite{IEEE1609-3}, the TXpower80211 value could be the EIRP in dBm, indicating the maximum EIRP power level that is allowed. 
A significant number of values are only available for 802.11-based communication technologies. 
In order to accommodate a wider range of technologies, enhancement will be necessary. 
Additionally, other values, such as latitude, longitude, and elevation, lack the requisite unit. 
In addition, the final three values have been redefined. 
A definition can be found in the ASN.1-ITS-Container by ETSI. 
Once again, the \cite{SAEJ2735} could be used.

The majority of new standards are only compatible with IPv6.

The service announcement delineates a limited number of communication-related parameters. 
The lack of precision in these parameters precludes the derivation of any meaningful information from them.

An extension to this model would be that stations that wish to consume a service can broadcast a service request message (SERM) to request a specific service. 
Stations that are able to provide such a service or the requested information can then start sending information with service-related information. 
This concept is known as the pull paradigm \cite{SAC2013}. 
However, this is not yet a standardized approach. 
In the ISO standard \cite{TS16460}, a service response message is specified, but this was not adopted by SI.


\section{Outlook}
\label{Chapter_ITSProtocols-outlook}
This white paper provides a comprehensive overview of the protocols utilized in the various ITS ecosystems.
It covers protocols from the application layer to the physical layer of the ISO reference model.
The variety of protocols employed in the different ecosystems that have emerged over the past decades presents challenges for interoperability and system interaction, which are essential for achieving the goal of safe and environmentally friendly future mobility.
The detailed analysis of traffic-related communication protocols reveals that the communication requirements are not entirely clear and require further elaboration.

\appendices
\section*{Acknowledgment}
This work was funded by the German projects ConnRAD (grant number 16KISR032) and Gaia-X4MoveID (grant number 19S22002L).

\ifCLASSOPTIONcaptionsoff
  \newpage
\fi

\bibliographystyle{IEEEtran}
%
\bibliography{bibtex/bib/refs}

\begin{IEEEbiographynophoto}
{Jonas Vogt} is a senior researcher at the University of Applied Sciences Saarland - htw saar. He is the co-team leader of the ITS research group (FGVT) and coordinates the competence center for ``Future Transportation Society (FTS)`` at htw saar. He was involved in over 25 national and European research project in the area of connected and automated mobility with focus on communication networks in combination with traffic infrastructure requirements, human behaviour, socio-economical aspects and technology acceptance. His research focuses on communication architectures and protocol design for connected and automated mobility. Jonas is an expert of the ETSI STF 585 on Multi-channel Operation for Release 2 and member of IEEE. He is currently pursuing his Ph.D. at Rheinland-Pfälzische Technische Universität Kaiserslautern-Landau, Germany. 
\end{IEEEbiographynophoto}

\vfill

\end{document}